\documentclass[10pt,pra,twocolumn,superscriptaddress,showpacs,floatfix]{revtex4-2}
\usepackage{graphics}
\usepackage{graphicx}
\usepackage{bm}
\usepackage{amsmath}
\usepackage{amsfonts}
\usepackage{amssymb}
\usepackage{latexsym}
\usepackage[dvipsnames]{xcolor}
\usepackage{hyperref}
\usepackage{float}
\begin{document}

\title{Cavity QED beyond the Jaynes-Cummings model}

\author{Abeer Al Ghamdi} 
\affiliation{Physics \& Astronomy Department, King Saud University, Riyadh 11362, Saudi Arabia.}
\affiliation{School of Physics and Astronomy, University of Leeds, Leeds LS2 9JT, United Kingdom.}
\author{Gin Jose} 
\affiliation{School of Chemical and Process Engineering, University of Leeds, Leeds LS2 9JT, United Kingdom.}
\author{Almut Beige}
\affiliation{School of Physics and Astronomy, University of Leeds, Leeds LS2 9JT, United Kingdom.}

\date{\today}

\begin{abstract}
As atom-cavity systems are becoming more sophisticated, the limitations of the Jaynes-Cummings model are becoming more apparent. In this paper, we therefore take a more dynamical approach to the modelling of atom-cavity systems and do not reduce the electromagnetic field inside the resonator to a single mode. Our approach shows that the decay rate $\Gamma_{\rm cav}$ of an emitter inside a subwavelength cavity with metallic mirrors can be much larger than its free space decay rate $\Gamma_{\rm free}$ due to constructive interference effects of the emitted light. In general, however, we find that $\Gamma_{\rm cav} = \Gamma_{\rm free}$ to a very good approximation which might explain why atom-cavity experiments with planar mirrors have not been able to operate in the so-called strong coupling regime. 
\end{abstract} 

\maketitle

\section{Introduction} \label{Intro}

\begin{figure}[t]
\centering
\includegraphics[width=0.9 \columnwidth]{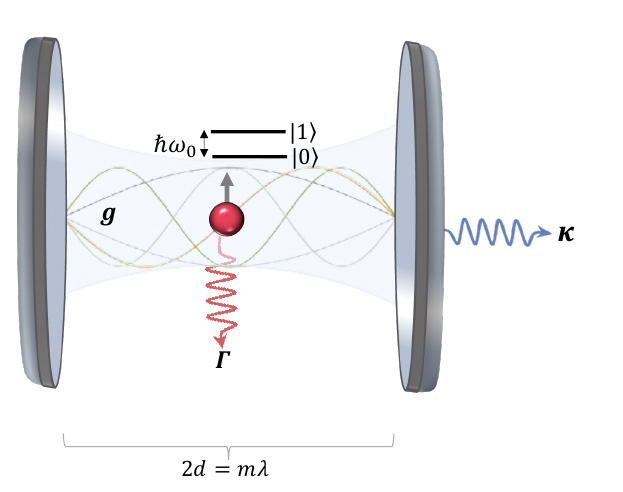}
\caption{[Colour online] Standard view of atom-cavity systems in quantum optics, as promoted by the Jaynes-Cummings Hamiltonian in Eq.~(\ref{I1}) and the corresponding master equations in Eq.~(\ref{I2}). When placed inside an optical cavity with a matching cavity frequency, the atom couples resonantly to a single standing wave mode of the quantised electromagnetic field and exchanges energy with the mode in a controlled way. Moreover, there are two decay channels: photons might leak out through the resonator mirrors with a decay rate $\kappa$ or might come directly from the atom which has the spontaneous decay rate $\Gamma$.}
\label{figpaperlogo2}
\end{figure}

The Jaynes-Cummings Hamiltonian for the modelling of light-matter interactions at the single atom level was first introduced in 1963 \cite{Jaynes, Shore,Jaynes2}. Since then it has become a cornerstone of quantum optics and has been used extensively to design and analyse atom-cavity experiments (cf.~e.g.~Refs.~\cite{Haroche,Rempe,Meschede1,Walther,Kimble}). Our standard approach to optical cavities is to assume that the mirrors impose strict boundary conditions on the quantised electromagnetic field. These reduce its possible excitations from a continuum of travelling photon modes effectively to a single standing wave field mode. When a single two-level atom with ground state $|0_{\rm A} \rangle$ and excited state $|1_{\rm A} \rangle$ is placed into an optical cavity, it interacts with this mode as described by the Jaynes-Cummings Hamiltonian 
\begin{eqnarray} \label{I1}
H_{\rm JC} &=& \hbar g \, \sigma^- a^\dagger + {\rm H.c.} 
\end{eqnarray}
As illustrated in Fig.~\ref{figpaperlogo2}, $g$ denotes the atom-cavity coupling constant, $a$ with $[a,a^\dagger] =1$ is the annihilation operator of a single cavity photon and $\sigma^- = |0_{\rm A} \rangle \langle 1_{\rm A}|$ is the atomic lowering operator. This interaction Hamiltonian conserves the number of excitations in the atom-cavity system. It also conserves the free energy of atom and field, as it should, since it commutes with the free Hamiltonian \cite{Stokes2}.

Quantum opticians usually assume that atom-cavity systems have two independent decay channels. When excited, the atom might spontaneously emit a photon into free space. Moreover, photons might spontaneously leak out through the resonator mirrors. Denoting the corresponding decay rates by $\Gamma$ and $\kappa$, respectively, it is assumed that the density matrix $\rho_{\rm AC}$ of atom and cavity evolves according to the master equation \cite{Agarwal,Breuer} 
\begin{eqnarray} \label{I2}
\dot \rho_{\rm AC} &=& - {{\rm i} \over \hbar} \left[ H_{\rm JC} , \rho_{\rm AC}  \right] \notag \\
&& - {\Gamma \over 2}  \left( 2 \sigma^- \rho_{\rm AC} \sigma^+ - \sigma^+ \sigma^- \rho_{\rm AC} - \rho_{\rm AC} \sigma^+ \sigma^- \right)  \notag \\
&& - {\kappa \over 2} \left( 2 a \rho_{\rm AC} a^\dagger - a^\dagger a \rho_{\rm AC} - \rho_{\rm AC} a^\dagger a \right) \, .
\end{eqnarray}
Notice that different from the master equations of individual atoms in free space \cite{Hegerfeldt,Stokes}, this equation has not been derived from basic principles. It is a phenomenological equation which preserves the trace and the positivity of the density matrix $\rho_{\rm AC}$ and simplifies to the master equation of a single atom when $\kappa = 0$. The above model of atom-cavity systems is widely known as the Jaynes-Cummings model \cite{theory,theory2,theory3,theory4}.  


In recent years, due to technological advances in nanofabrication, experimental realisations of atom-cavity systems have become more and more sophisticated \cite{Jorgensen2025}. In addition, to traditional optical cavities, scientists now perform experiments with individual atoms and quantum dots coupled to ring resonators \cite{Jelena}, to plasmonic subwavelength cavities \cite{Pelton,Baumberg,Baumberg2,Russell2012,Li2024}, to photonic crystal cavities \cite{Yablonovitch1987,Jelena2,Lohdahl2004,Krishnamoorthy2010} 
and to fiber tip cavities \cite{Kuhn,Reichel}. In addition, cavities mirrors can be realised using metamaterials \cite{complex,Barreda2021,Liu2025}. Optical cavities might even be realised with the help of mirror arrays\textemdash so-called Bragg stacks\textemdash in which case it becomes difficult to identify the resonance frequency and the effective length of the resonator \cite{Guerin}. For many of these systems, the relatively simple Jaynes-Cummings model is no longer applicable \cite{Barlow,Kuhnpush,Liberato,Arno}. 


More concretely, Blaha {\em et al.}~\cite{Arno} recently pointed out that the Jaynes–Cummings model breaks down when an atom is placed inside an optical cavity in relatively close vicinity of one of the resonator mirrors. In this case, atom-mirror interactions lead to interference effects which can alter the emission rates. Traditional models also struggle to predict the strongly enhanced spontaneous decay rates of individual emitters in plasmonic subwavelength cavities \cite{Pelton,Baumberg,Baumberg2,Russell2012,Li2024}. Since the distance of the mirrors of these subwavelength cavities is much smaller than the wavelength associated with the transition frequency $\omega_0$ of the atom, which is typically in the optical regime, there is no resonant standing-wave cavity field mode that the atom could couple to. It therefore came as a surprise that the spontaneous decay of an atom passing through such a resonator is significantly enhanced. These findings challenge existing theoretical paradigms and suggest that light-matter interactions are still not yet fully understood \cite{Ozdemir2017, Krishnamoorthy2017, Baumberg2}.


Another example, which challenges the usual single-mode description of the quantised electromagnetic field between two mirrors, are so-called Fabry-Perot cavities \cite{Abeer}. Such cavities are well understood in classical optics \cite{Hecht} and have a very high transmission rate for resonant incoming light, while their transmission rate for off-resonant light is very low. In a quantum optical model involving only a single standing-wave resonator mode, it is impossible to encode the information whether the laser excites the resonator from the left or from the right. Hence, since the Jaynes-Cummings model is symmetric with respect to exchanging both cavity mirrors, no predictions can be made through which mirror the photons  eventually leave the resonator. To fix this problem, two types of photons, namely left- {\em and} right-moving photons, need to be considered to capture the typical behaviour of Fabry-Perot cavities \cite{Barlow}.


Motivated by the above observations, the presence of cavity mirrors is in the following taken into account by considering how they affect the dynamics of local field excitations and not by reducing the size of the Hilbert space of the quantised electromagnetic field inside the resonator. In other words, we assume here that the quantised electromagnetic field is the same in the presence and in the absence of cavity mirrors. The only thing that changes is the field Hamiltonian \cite{Jake,AMC}. In addition, we assume that the emitter is much smaller than the wavelength of the emitted light, so that the re-absorption of photons is essentially negligible. As we have already demonstrated in Refs.~\cite{Nick,Dawson}, all we need to know in this case to derive the master equation of a single atom in the presence of optical elements, like resonator mirrors, is how local electric field vectors evolve in the {\em absence} of atom-field interactions. As we shall see below, a single atom between two parallel plates evolves to a very good approximation like an atom in free space, i.e.~according to a master equation of the form
\begin{eqnarray} \label{I3}
\dot \rho_{\rm A} &=& {\Gamma_{\rm cav} \over 2}  \left( 2 \sigma^- \rho_{\rm A} \sigma^+ - \sigma^+ \sigma^- \rho_{\rm A} - \rho_{\rm A} \sigma^+ \sigma^- \right) \, .
\end{eqnarray}
However, due to interference effects, the spontaneous decay rate $\Gamma_{\rm cav}$ in this equation differs in general from the free space decay rate $\Gamma_{\rm free}$ of the emitter. Notice that there is a difference between Eqs.~(\ref{I2}) and (\ref{I3}). They cannot both be valid at the same time, unless the cooperativity parameter $C= g^2/\kappa \Gamma$ is much smaller than one (weak coupling regime). In this case, Eq.~(\ref{I2}) can be transformed into Eq.~(\ref{I3}) via an adiabatic elimination of the cavity mode. 


To test the validity of our approach to the modelling of atom-cavity system, we first derive the quantum optical master equations of a single atom in free space and in the presence of a partially transparent mirror in an analogous way. The study of spontaneous photon emission near reflecting surfaces has a long history \cite{Purcell1946,Drexhage,Kleppner1981,Chance1978,Dorner,Kuraptsev}. In addition, experiments were performed such as those by Drexhage \cite{Drexhage1974} in the 1970s, by Amos and Barnes in the 90-ies \cite{Amos1997}, and by Eschner et al.~\cite{Eschner} in the 1990's. As we shall see below, the dependence of our predicted spontaneous decay rates on the atom-mirror distance is in agreement with these experiments for certain mirror reflection rates. Afterwards, we use the same methodology to obtain an expression for $\Gamma_{\rm cav}$ in Eq.~(\ref{I3}). Depending on the distance $d$ of the resonator mirrors, we distinguish in the following between optical and subwavelength cavities.


As part of our theoretical approach, certain simplifications were introduced to render the problem analytically tractable. These are discussed in detail in the conclusion of the manuscript. For example, we assume in the following that the emitter has been placed in the middle between two parallel planar mirrors with finite reflection and transmission rates, as illustrated in Fig.~\ref{figpaperlogo2}. Moreover, we assume that local field excitations, once they have left the emitter, propagate as they would in the absence of atom-field interactions without ever being re-absorbed by the emitter. This approximation is well justified as long as the atomic source is much smaller than the wavelength of the emitted light. Usually, experimentalists encourage the re-absorption of light by carefully shaping mirror surfaces \cite{Kuhn,Reichel} or by offering a whole stack of mirror surfaces \cite{Guerin}. Nevertheless, we obtain some interesting results. 


In general, we find that the spontaneous decay rate $\Gamma_{\rm cav}$ of an emitter inside an {\em optical} resonator depends neither on the distance between the cavity mirrors $d$ nor their reflection and transmission rates. It also does not depend on the transition frequency $\omega_0$ of the atom and equals instead the free space emission rates $\Gamma_{\rm free}$ to a very good approximation. This observation is not surprising, if one takes into account that the spontaneous decay rate of an atom in the presence of a partially transparent mirror also equals $\Gamma_{\rm free}$ unless the atom-mirror distance is of the order of the wavelength $\lambda_0$ of the emitted light. Our equations confirm that it is difficult to enhance atom-cavity interactions by simply reducing the distance between the cavity mirrors or by increasing their quality. Operating experiments in the so-called strong coupling regime requires additional tricks, like the careful shaping of cavity mirrors in order to encourage the re-absorption of light by the atom \cite{Reichel,Kuhn}. 


An exception are {\em plasmonic subwavelength} cavities, i.e.~cavities with mirrors whose distance $d$ is much smaller than the transition wavelength $\lambda_0$ of the emitter. It is shown in this paper that the decay rate $\Gamma_{\rm cav}$ inside a subwavelength cavity can exceed $\Gamma_{\rm free}$ by several orders of magnitude, if electric field amplitudes of light do not accumulate a minus sign upon reflection by the cavity mirrors. In addition, the resonator mirrors have to be highly-reflecting. This observation is in good agreement with experimental observations \cite{Pelton,Russell2012,Baumberg,Baumberg2,Li2024} which utilise plasmonic nanocavities and demonstrate an enhanced photon emission from quantum dots and nitrogen-vacancy centres. Usually, this enhancement is attributed to extremely high Purcell factors due to strong spatial confinement of light and a high density of optical states within the cavity \cite{Li2024}. For example, it has been reported that plasmonic nanocavities\textemdash where light is confined to volumes several orders of magnitude smaller than the diffraction limit\textemdash can theoretically achieve Purcell factors exceeding 10,000 \cite{Baumberg,Russell2012}. Here we attribute the enhanced emission instead to constructive interference effects in the far-field of the emitted light. 


Our calculations also show that similar enhancement of $\Gamma_{\rm cav}$ cannot be achieved in photonic bandgap materials \cite{Yablonovitch1987,Jelena2,Lohdahl2004,Krishnamoorthy2010}. In contrast to plasmonic mirrors, like metals and metasurfaces, which enable surface currents and consequently exhibit complex reflection rates, the reflection rates of photonic bandgap materials are always real and negative \cite{Fox}. The reason for this is that the resonator mirrors are made of dielectric media in this case. Upon light reflection by a dielectric medium, electric field amplitudes accumulate a minus sign to ensure that they remain continuous across the mirror surface \cite{Hecht,AMC}. Usually, the lack of emission enhancement is assumed to be due to significant non-radiative losses \cite{Russell2012} and the absence of available photonic modes. Here we find instead that the lack of enhancement of light emission is due to a lack of constructive interference effects.


This paper is organised as follows. In Section~\ref{sec2}, we review the derivation of the quantum optical master equation in Eq.~(\ref{I3}) and obtain an analytical expression for $\Gamma_{\rm cav}$. Afterwards, we analyse this expression for the case of an emitter in different structured environments. Section~\ref{sec3} focuses on the dynamics of an atom in free space and in the presence of a partially-transparent mirror surface. In Section~\ref{sec4}, we calculate the spontaneous decay $\Gamma_{\rm cav}$ of a single emitter inside a subwavelength and an optical cavity with positive and negative mirror reflection rates. Finally, in Section~\ref{sec5}, we summarize our results and discuss their implications.

\section{The spontaneous decay rate of a single atom} \label{sec2}

In this section, we review the derivation of the quantum optical master equations of a single atom with ground state $|0_{\rm A} \rangle $ and excited state $|1_{\rm A} \rangle $ following an approach that was first introduced in Ref.~\cite{Dawson}. These equations describe the dynamics of the atomic density matrix $\rho_{\rm A} (t) $. As we shall see below, the time derivative of $\rho_{\rm A} (t) $ can always be written as 
\begin{eqnarray} \label{ME}
\dot \rho_{\rm A} = -\frac{\rm i}{\hbar} \left[ H_{\rm cond} \rho_{\rm A} - \rho_{\rm A} H_{\rm cond}^\dagger \right]+\mathcal{L}(\rho_{\rm A}) \, . 
\end{eqnarray}
Here $H_{\rm cond}$ and $\mathcal{L}(\rho_{\rm A})$ represent a non-Hermitian Hamiltonian and a reset operator which describes the unnormalised state of the atom immediately after an emission. In the remainder of this section, we show that $\mathcal{L}(\rho_{\rm A})$ and $H_{\rm cond} $ are of the general form
\begin{eqnarray} \label{LH}
\mathcal{L}(\rho_{\rm A}) &=& \Gamma \, \sigma^- \rho_{\rm A} \sigma^+ \, , \notag \\
H_{\rm cond} &=& H_{\rm A} - \frac{\rm i}{2} \hbar \, \Gamma \, \sigma^+  \sigma^- \, ,
\end{eqnarray}
if we neglect a level shift of the excited atomic state \cite{Hegerfeldt,Stokes}. Here $H_{\rm A}$ and $\Gamma$ denote the Hamiltonian and  the spontaneous decay rate of the atom and $\sigma^- = |0_{\rm A} \rangle \langle 1_{\rm A}|$ and $\sigma^+ = |1_{\rm A} \rangle \langle 0_{\rm A}|$ are lowering and raising operators. In this section, we derive an expression which allows us to calculate $\Gamma$ later on in different environments.

\subsection{The Hamiltonian of atom and field}

In the Schr\"odinger picture, the total Hamiltonian $H$ of atom and field can be written as
\begin{eqnarray} \label{Htotal}
H &=& H_{\rm A} + H_{\rm AF} + H_{\rm F} 
\end{eqnarray}
with $H_{\rm F}$ and $H_{\rm AF}$ denoting the Hamiltonian of the free radiation field and the atom-field interaction term. Using the above notation, $H_{\rm A}$ is given by
\begin{eqnarray} \label{HA}
 H_{\rm A} &=& \hbar \omega_0 \, \sigma^+  \sigma^- 
\end{eqnarray}
where $\omega_0$ represents the transition frequency of the atom. Within the usual dipole and the rotating wave approximation the atom-field interaction Hamiltonian $H_{\rm AF}$ moreover equals \cite{theory,theory2,theory3,theory4,Stokes}
\begin{eqnarray} \label{8}
 H_{\rm AF} &=& - e \, {\boldsymbol {\cal D}}_{01} \, \sigma^- \cdot \boldsymbol{\cal E}^\dagger (\boldsymbol{r}_0) + {\rm H.c.} 
  \end{eqnarray}
Here $e$ is the charge of a single electron and  ${\boldsymbol {\cal D}}_{01}$ represents the complex dipole moment of the atom. Moreover, $\boldsymbol{\cal E}^\dagger (\boldsymbol{r}_0)$ represents the complex electric field observable at the position $\boldsymbol{r}_0$ of the atom. For simplicity, we assume in the following that the vector ${\boldsymbol {\cal D}}_{01}$ is real and given by
\begin{eqnarray} \label{UFx}
{\boldsymbol {\cal D}}_{01} &=& \|{\boldsymbol {\cal D}}_{01}\| \, (0,0,1) \, .
\end{eqnarray}
As we shall see below, we do not require an explicit expression for the field Hamiltonian $H_{\rm F}$ in Eq.~(\ref{Htotal}), as long as we know how local field excitations created at the position of the atom evolve in time.

As in Refs.~\cite{Nick,Dawson}, we write the field observable $\boldsymbol{\cal E}^\dagger (\boldsymbol{r})$ at position $\boldsymbol{r}$ in the following as the sum over different $(\boldsymbol{s},\lambda)$ contributions, 
\begin{eqnarray} \label{E}
\boldsymbol{\cal E}^\dagger (\boldsymbol{r}) &=& \sum_{\lambda = {\sf H}, {\sf V}} \int_{\cal S} {\rm d}^2 s \, \boldsymbol{\cal E}^\dagger_{\boldsymbol{s} \lambda} (\boldsymbol{r}) \, , 
\end{eqnarray}
with $\lambda = {\sf H}, {\sf V}$ denoting  the polarisations of the corresponding electric field vectors and with the unit vectors $\boldsymbol{s} \in {\cal S}$ indicating directions of propagation. Each $\boldsymbol{\cal E}^\dagger_{\boldsymbol{s} \lambda} (\boldsymbol{r}) $ component can be decomposed further into electric field contributions with wave vectors $\boldsymbol{k} = k \boldsymbol{s}$, unit polarisation vector ${\boldsymbol e}_{\boldsymbol{s} \lambda}$ and photon creation operators $a^\dagger_{\boldsymbol{k} \lambda}$ such that \cite{Bennett}
\begin{eqnarray} \label{Esl}
\boldsymbol{\cal E}^\dagger_{\boldsymbol{s} \lambda}(\boldsymbol{r}) 
&=& \int_0^\infty {\rm d}{k} \,  k^2 \, \sqrt{\frac{\hbar ck}{4 \pi^3 \varepsilon_0}} \, {\rm e}^{{\rm i} {\boldsymbol k} \cdot {\boldsymbol r}} \, a^\dagger_{\boldsymbol{k} \lambda} \, {\boldsymbol e}_{\boldsymbol{s} \lambda} \, . ~~
\end{eqnarray}
Here $\varepsilon_0$ denotes the permittivity of the respective medium. The $a^\dagger_{\boldsymbol{k} \lambda}$ operators and the corresponding annihilation operators $a_{\boldsymbol{k} \lambda}$ obey the bosonic commutator relation
\begin{eqnarray} \label{Esl2}
\big[ a_{\boldsymbol{k}' \lambda'}, a^{\dagger}_{\boldsymbol{k} \lambda} \big] 
&=& \delta_{\lambda \lambda'} \, \delta^3 (\boldsymbol{k} - \boldsymbol{k}') \notag \\
&=& {1 \over k^2} \, \delta_{\lambda \lambda'} \, \delta^2 (\boldsymbol{s} - \boldsymbol{s}') \, \delta(k-k')
\end{eqnarray}
with $\boldsymbol{k} = k \boldsymbol{s}$ and ${\boldsymbol e}_{\boldsymbol{s} \lambda}\cdot \boldsymbol{s}=0$, while ${\boldsymbol e}_{\boldsymbol{s} \lambda} \cdot {\boldsymbol e}_{\boldsymbol{s} \lambda'} = \delta_{\lambda\lambda'}$. For later convenience, we now use the above equations to calculate the scalar product $\langle 0_{\rm}| {\boldsymbol {\cal D}}_{01} \cdot  \boldsymbol{\cal E}_{\boldsymbol{s}' \lambda'}(\boldsymbol{r}') ~ {\boldsymbol {\cal D}}_{01} \cdot  \boldsymbol{\cal E}^\dagger_{\boldsymbol{s} \lambda}(\boldsymbol{r}) |0_{\rm F} \rangle$ where $|0_{\rm F} \rangle$ represents the vacuum state of the free radiation field. This expression equals
\begin{widetext}
\begin{eqnarray} \label{Esl22}
\langle 0_{\rm F}| {\boldsymbol {\cal D}}_{01} \cdot \boldsymbol{\cal E}_{\boldsymbol{s}' \lambda'}(\boldsymbol{r}') ~ {\boldsymbol {\cal D}}_{01} \cdot \boldsymbol{\cal E}^\dagger_{\boldsymbol{s} \lambda}(\boldsymbol{r}) |0_{\rm F} \rangle 
&=& \frac{\hbar c}{4 \pi^3 \varepsilon_0} \, \delta_{\lambda \lambda'} \, \delta^2 (\boldsymbol{s} - \boldsymbol{s}') 
\int_0^\infty {\rm d}{k} \int_0^\infty {\rm d}{k'} \sqrt{kk'^5} \, \delta(k-k') \, {\rm e}^{{\rm i} k ({\boldsymbol s} \cdot {\boldsymbol r} - {\boldsymbol s}' \cdot {\boldsymbol r}')} \, \big\| {\boldsymbol {\cal D}}_{01} \cdot \boldsymbol{e}_{{\boldsymbol s}\lambda} \big\|^2 \notag \\
&=& \frac{\hbar c}{4 \pi^3 \varepsilon_0} \, \delta_{\lambda \lambda'} \, \delta^2 (\boldsymbol{s} - \boldsymbol{s}') 
\int_0^\infty {\rm d}{k} \,  k^3 \, {\rm e}^{{\rm i} k ({\boldsymbol s} \cdot {\boldsymbol r} - {\boldsymbol s}' \cdot {\boldsymbol r}')} \, \big\| {\boldsymbol {\cal D}}_{01} \cdot \boldsymbol{e}_{{\boldsymbol s}\lambda} \big\|^2 \notag \\
&=& \frac{{\rm i} \hbar c}{8 \pi^2 \varepsilon_0} \, \delta_{\lambda \lambda'} \, \delta^2 (\boldsymbol{s} - \boldsymbol{s}') \, \delta^{(3)}( {\boldsymbol s} \cdot {\boldsymbol r} - {\boldsymbol s}' \cdot {\boldsymbol r}') \, \big\| {\boldsymbol {\cal D}}_{01} \cdot \boldsymbol{e}_{{\boldsymbol s}\lambda} \big\|^2 \, . ~~
\end{eqnarray}
\end{widetext}
The function $\delta^{(3)}(x)$ in this equation denotes the third derivative of the Dirac delta function $\delta(x)$ with respect to $x$. The field Hamiltonian $H_{\rm F}$ depends on whether we consider an atom in free space, in the presence of a mirror interface or inside an optical cavity \cite{AMC}.

\subsection{Derivation of $\mathcal{L}(\rho_{\rm A})$ and $H_{\rm cond}$}

In the following, we assume that the atom is always surrounded by a free radiation field in its vacuum state, since the atom is the only light source present and any previously generated field excitations travel away at the speed of light $c$ and can no longer excite the atom. Suppose the density matrix of the atom initially equals $\rho_{\rm A}(0)$. Then the total density matrix $\rho(0)$ of atom and field is given by
\begin{eqnarray} \label{10}
\rho(0) &=& |0_{\rm F} \rangle \, \rho_{\rm A}(0) \, \langle 0_{\rm F}| \, . 
\end{eqnarray}
Evolving this density matrix for a time $\Delta t$ with the time evolution operator $U(\Delta t,0)$ associated with the Hamiltonian $H$ in Eq.~(\ref{Htotal}) and then taking the trace over the field degrees of freedom reveals the atomic density matrix 
\begin{eqnarray} \label{11}
\rho_{\rm A} (\Delta t) &=& {\rm Tr}_{\rm F} \left( U(\Delta t,0) \, \rho(0) \, U^\dagger(\Delta t,0) \right) . ~~
\end{eqnarray}
Here the trace over the field is taken, since we are only interested in how the state of the atom evolves in time and ignore the dynamics of field excitations. To determine the derivative of the atomic density matrix $\rho_{\rm A}(0)$, we now calculate $\rho_{\rm A} (\Delta t)$ up to first order in $\Delta t$ and consider the expression
\begin{eqnarray} \label{12}
\dot \rho_{\rm A}(0) &=& \frac{1}{\Delta t} \left(\rho_{\rm A}(\Delta t)-\rho_{\rm A}(0)\right)
\end{eqnarray}
in the limit of small $\Delta t$. However, $\Delta t$ should not be too short such that the atom has enough time to transfer some of its energy into the radiation field. 

Next we notice that the atom-field interaction term $H_{\rm AF}$ only constitutes a small perturbation compared to the free Hamiltonian $H_0 = H_{\rm A} + H_{\rm F}$. In order to be able to take this into account, we change temporarily into the interaction picture with respect to $t=0$ and $H_0$. In the interaction picture, the Hamiltonian of atom and field equals
\begin{eqnarray} \label{Htotal2}
H_{\rm I}(t) &=& U_0^\dagger (t,0) \, H_{\rm AF} \, U_0(t,0)
\end{eqnarray}
where $U_0(t,0)$ denotes the time evolution operator of the free dynamics. Here $U_{\rm A}(t,0)$ and $U_{\rm F}(t,0)$ are the time evolution operators corresponding to $H_{\rm A}$ and $H_{\rm F}$, respectively. Since both Hamiltonians commute, we have $U_0 (t,0) =U_{\rm A}(t,0) \, U_{\rm F}(t,0)$. As we shall see below, it is advantageous to separate the free evolution of the field excitations from all remaining operators. We therefore also introduce the short hand notation 
\begin{eqnarray} \label{DDR3}
H_{\rm AF}(t) = U^\dagger_{\rm A}(t,0) \, H_{\rm AF} \, U_{\rm A}(t,0)
\end{eqnarray}
which allows us to write $H_{\rm I}(t)$ as
\begin{eqnarray} \label{DDR}
H_{\rm I}(t) &=& U^\dagger_{\rm F}(t,0) \, H_{\rm AF}(t) \, U_{\rm F}(t,0) \, .
\end{eqnarray}
From Eqs.~(\ref{HA}) and (\ref{8}), we conclude that
\begin{eqnarray} \label{DD2}
H_{\rm AF}(t) &=& - e \, {\rm e}^{-{\rm i} \omega_0 t} \, {\boldsymbol {\cal D}}_{01} \cdot \boldsymbol{\cal E}^\dagger (\boldsymbol{r}_0) \, \sigma^- + {\rm H.c.} 
\end{eqnarray}
with ${\boldsymbol {\cal D}}_{01}$ given in Eq.~(\ref{UFx}). The above equations allow us to write the density matrix $\rho_{\rm A}(\Delta t)$ in the following in a relatively compact form.

Using second-order perturbation theory, we can write the time evolution operator $U_{\rm I}(\Delta t,0)$ in the interaction picture as
\begin{eqnarray} \label{BB3}
U_{\rm I}(\Delta t,0) &=& 1 - {{\rm i} \over \hbar} \int_0^{\Delta t} {\rm d}t \, H_{\rm I}(t) \notag \\
&& - {1 \over \hbar^2} \int_0^{\Delta t} {\rm d}t \int_0^{t} {\rm d}t' \, H_{\rm I}(t) H_{\rm I}(t') \, . 
\end{eqnarray}
Moreover, $U(\Delta t,0) $ in the Schr\"odinger picture equals
\begin{eqnarray} \label{BB40x}
U(\Delta t,0) &=& U_0 (\Delta t,0) \, U_{\rm I}(\Delta t,0) \notag \\
&=& U_{\rm F} (\Delta t,0) \, U_{\rm A} (\Delta t,0) \, U_{\rm I}(\Delta t,0) \, .
\end{eqnarray}
In addition, we take into account that $H_{\rm I}(t)$ either creates a photon and de-ecxites the atom or annihilates a photon and excites the atom.  When combining Eq.~(\ref{11}) with Eqs.~(\ref{DDR})-(\ref{BB40x}), we can therefore now show that
\begin{widetext} 
\begin{eqnarray} \label{BB4000}
\rho_{\rm A}(\Delta t) &=& U_{\rm A} (\Delta t,0) \, \rho_{\rm A}(0) \,  U^\dagger_{\rm A} (\Delta t,0) \notag \\
&& + \frac{e^2}{\hbar^{2}} \int_{0}^{\Delta t} {\rm d}t \int_0^{\Delta t} {\rm d}t' \, {\rm e}^{{\rm i} \omega_0 (t-t')} \, 
\langle 0_{\rm F}| \, {\boldsymbol {\cal D}}_{01} \cdot \boldsymbol{\cal E} (\boldsymbol{r}_0) \, U_{\rm F} (t,t') \, {\boldsymbol {\cal D}}_{01} \cdot \boldsymbol{\cal E}^\dagger (\boldsymbol{r}_0) |0_{\rm F} \rangle \, \sigma^- \, \rho_{\rm A}(0) \, \sigma^+ \nonumber \\
&& - \frac{e^2}{\hbar^2} \int_0^{\Delta t} {\rm d}t \int_0^{t} {\rm d}t' \, {\rm e}^{{\rm i} \omega_0 (t-t')} \, \langle 0_{\rm F}| {\boldsymbol {\cal D}}_{01} \cdot \boldsymbol{\cal E} (\boldsymbol{r}_0) \, U_{\rm F} (t,t') \,  {\boldsymbol {\cal D}}_{01} \cdot \boldsymbol{\cal E}^\dagger (\boldsymbol{r}_0) |0_{\rm F} \rangle  \, \sigma^+ \sigma^- \, \rho_{\rm A}(0) + {\rm H.c.}
\end{eqnarray}
in first order in $\Delta t$. In addition, we know from Eq.~(\ref{ME}) that
\begin{eqnarray} \label{ME2}
\rho_{\rm A} (\Delta t) &=& \rho_{\rm A}(0) - \frac{\rm i}{\hbar} \left[ H_{\rm cond} \rho_{\rm A}(0) - \rho_{\rm A}(0) H_{\rm cond}^\dagger \right] \Delta t + \mathcal{L}(\rho_{\rm A}(0)) \, \Delta t \, .
\end{eqnarray}
Hence when comparing Eqs.~(\ref{BB4000}) and (\ref{ME2}), we find that  
\begin{eqnarray} \label{BB4}
{\cal L} (\rho_{\rm A}(0)) &=& \frac{e^2}{\hbar^{2} \Delta t} 
\int_{0}^{\Delta t} {\rm d}t \int_0^{\Delta t} {\rm d}t' \, {\rm e}^{{\rm i} \omega_0 (t-t')} \, \langle 0_{\rm F}| {\boldsymbol {\cal D}}_{01} \cdot \boldsymbol{\cal E} (\boldsymbol{r}_0) \, U_{\rm F} (t,t') \,  {\boldsymbol {\cal D}}_{01} \cdot \boldsymbol{\cal E}^\dagger (\boldsymbol{r}_0) |0_{\rm F} \rangle \, \sigma^- \, \rho_{\rm A}(0) \, \sigma^+ \, , \nonumber \\
H_{\rm cond} &=& H_{\rm A} - \frac{{\rm i}e^2}{\hbar \Delta t} \int_0^{\Delta t} {\rm d}t \int_0^{t} {\rm d}t' \,
 {\rm e}^{{\rm i} \omega_0 (t-t')} \, \langle 0_{\rm F}| {\boldsymbol {\cal D}}_{01} \cdot \boldsymbol{\cal E} (\boldsymbol{r}_0) \, U_{\rm F} (t,t') \,  {\boldsymbol {\cal D}}_{01} \cdot \boldsymbol{\cal E}^\dagger (\boldsymbol{r}_0) |0_{\rm F} \rangle \, \sigma^+ \sigma^- \, .  
\end{eqnarray}
\end{widetext}
The above equations show that all we need to know to calculate ${\cal L} (\rho_{\rm A}(0))$ and $H_{\rm cond}$ is how local electric field excitations created at the position $\boldsymbol{r}_0$ of the atom evolve in time in the absence of atom-field interactions. As we shall see in the next two sections, the dynamics of these local field excitations can be deduced from classical optics in a relatively straightforward way after decomposing the electric field observable $ \boldsymbol{\cal E}^\dagger (\boldsymbol{r}_0)$ as in Eq.~(\ref{E}) into its $({\boldsymbol s},\lambda)$ components.

\subsection{The spontaneous decay rate $\Gamma$}
 
To obtain an expression for the spontaneous decay rate $\Gamma$ which we introduced in Eq.~(\ref{LH}), we need to neglect an atomic level shift, i.e.~a term of the form $\hbar \Delta \, \sigma^+ \sigma^-$, in the expression for the conditional Hamiltonian. This level shift is in general much smaller than the atomic transition frequency $\omega_0$ and can therefore be absorbed into the energy $\hbar \omega_0$ of the excited state $|1_{\rm A} \rangle $ of the atom. To implement this approximation, we replace $H_{\rm cond}$ in the following by the operator $(H_{\rm cond} - {\rm H.c.})/2$. Subsequently we find that the above expressions coincide indeed with the reset operator $\mathcal{L}(\rho_{\rm A}(0))$ and with the conditional Hamiltonian $H_{\rm cond}$ in Eq.~(\ref{LH}), if we define the (real) spontaneous decay rate $\Gamma$ such that
\begin{eqnarray} \label{22}
\Gamma 
&=& \frac{e^2}{\hbar^2 \, \Delta t} \sum_{\lambda,\lambda' = {\sf H}, {\sf V}}  \int_0^{\Delta t} {\rm d}t \int_0^{\Delta t} {\rm d}t' \int_{\cal S} {\rm d}^2 s \int_{\cal S} {\rm d}^2 s' \notag \\
&& \times  {\rm e}^{{\rm i} \omega_0 (t-t')} \, 
\langle 0_{\rm F}| \, {\boldsymbol {\cal D}}_{01}  \cdot \boldsymbol{\cal E}_{\boldsymbol{s}' \lambda'} (\boldsymbol{r}_0) \, U^\dagger_{\rm F} (t',0) \notag \\
&& \hspace*{2.1cm} \times {\boldsymbol {\cal D}}_{01} \cdot U_{\rm F} (t,0) \, \boldsymbol{\cal E}_{\boldsymbol{s} \lambda}^\dagger (\boldsymbol{r}_0) |0_{\rm F} \rangle 
\end{eqnarray}
where Eq.~(\ref{E}) has been used. 

\section{A single atom in free space and in the presence of a partially transparent mirror} \label{sec31}

To check the validity of the above equations, we now use Eq.~(\ref{22}) to derive the well-known spontaneous decay rates $ \Gamma_{\rm free}$ and $\Gamma_{\rm mir}$ of a single two-level atom in free space \cite{Hegerfeldt,Stokes} and in the presence of a partially transparent mirror \cite{Nick,Dorner,Kuraptsev}, respectively. In both cases, our result agrees with the predictions of previous authors.
 
\subsection{Free space} \label{sec3}

\begin{figure*}[t]
\centering
\includegraphics[width=1.4 \columnwidth]{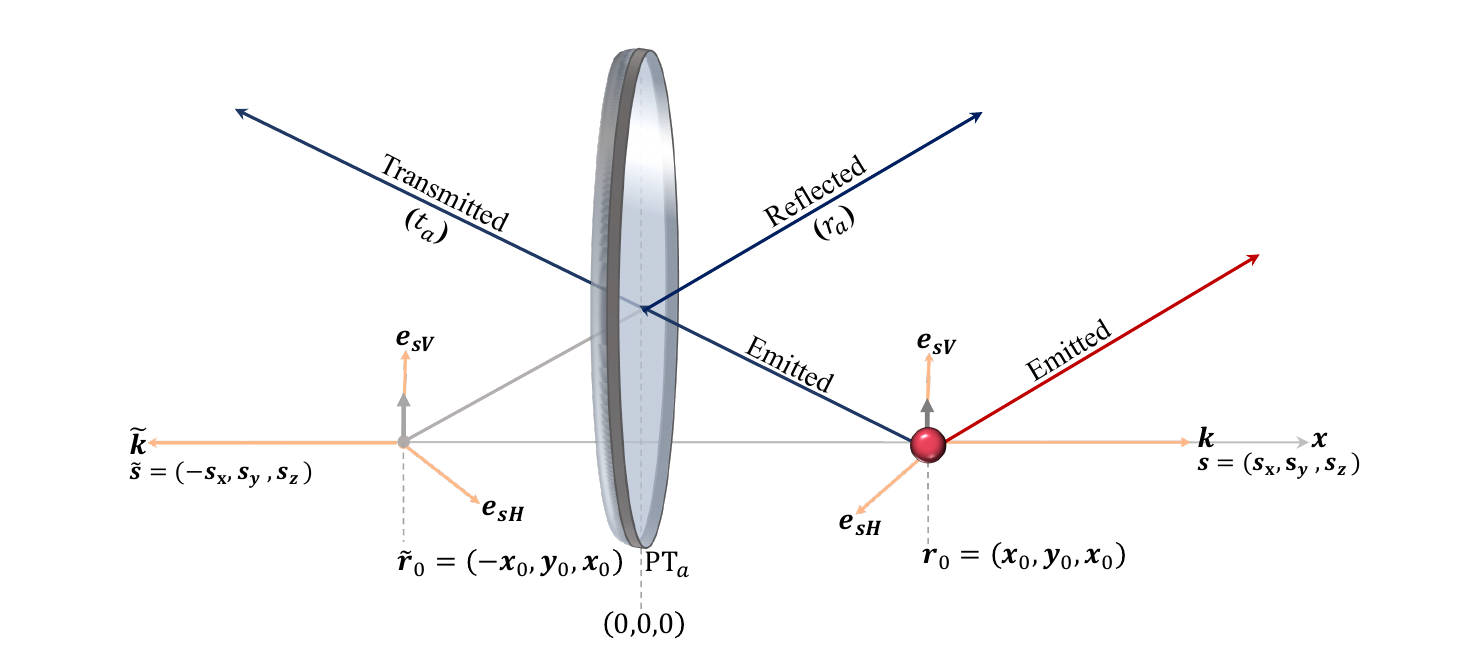}
\caption{[Colour online] Schematic view of an atom at a position $\boldsymbol{r}_0 = (x_0, 0, 0)$ with ${x_0>0}$ in the presence of a partially transparent mirror in the $x=0$ plane. Light emitted towards the mirror surface is  reflected, and seems to come from the position $\tilde {\boldsymbol r}_0 = (-x_0, 0, 0)$ of the mirror image of the atom and interferes with light emitted away from the mirror. For small atom-mirror distances, interference is mostly destructive and the spontaneous decay rate of the atom decreases. Here $ r_a$ and $t_a$ denote the (complex) reflection and transmission rates of the mirror interface.}
\label{figpaperlogo3}
\end{figure*}

From classical optics, we know that, in free space, light with a well defined direction ${\boldsymbol s}$ of propagation travels at constant speed along straight lines. Hence
\begin{eqnarray} \label{UF}
U_{\rm F} (t,0) \, \boldsymbol{\cal E}^\dagger_{\boldsymbol{s} \lambda}(\boldsymbol{r}_0) |0_{\rm F} \rangle
&=& \boldsymbol{\cal E}^\dagger_{\boldsymbol{s} \lambda}(\boldsymbol{r}_0 + \boldsymbol{s}ct) |0_{\rm F} \rangle \, .
\end{eqnarray}
Substituting this equation into Eq.~(\ref{22}), we therefore find that
\begin{eqnarray} \label{22new}
\Gamma_{\rm free} 
&=& \frac{e^2}{\hbar^2 \, \Delta t} \sum_{\lambda,\lambda' = {\sf H}, {\sf V}}  \int_0^{\Delta t} {\rm d}t \int_0^{\Delta t} {\rm d}t' \int_{\cal S} {\rm d}^2 s \int_{\cal S} {\rm d}^2 s' \notag \\
&& \times {\rm e}^{{\rm i} \omega_0 (t-t')} \, 
\langle 0_{\rm F}| \, {\boldsymbol {\cal D}}_{01}  \cdot \boldsymbol{\cal E}_{\boldsymbol{s}' \lambda'} (\boldsymbol{r}_0 + \boldsymbol{s}' ct') \notag \\
&&  \hspace*{2.1cm} \times {\boldsymbol {\cal D}}_{01} \cdot \boldsymbol{\cal E}_{\boldsymbol{s} \lambda}^\dagger (\boldsymbol{r}_0 + \boldsymbol{s} ct ) |0_{\rm F} \rangle \, .
\end{eqnarray}
Hence using Eq.~(\ref{Esl22}) leads us to
\begin{eqnarray} \label{22new}
\Gamma_{\rm free} 
&=& \frac{{\rm i} c e^2}{8 \pi^2 \varepsilon_0 \hbar \, \Delta t} \,  
\sum_{\lambda = {\sf H}, {\sf V}}  \int_0^{\Delta t} {\rm d}t \int_0^{\Delta t} {\rm d}t' \int_{\cal S} {\rm d}^2 s \notag \\
&& \times {\rm e}^{{\rm i} \omega_0 (t-t')} \, \delta^{(3)}(ct - ct') \, \big\| {\boldsymbol {\cal D}}_{01} \cdot \boldsymbol{e}_{{\boldsymbol s}\lambda} \big\|^2 \, .
\end{eqnarray}
To perform the time integrations in the above expression, we notice that
\begin{eqnarray} \label{help2}
\int_0^{\Delta t} \mathrm{d} t' \, \mathrm{e}^{- \mathrm{i} \omega_0 t'} \, \delta^{(3)}(ct-ct') &=& - {{\rm i} \omega_0^3 \over c^4} \, \mathrm{e}^{- \mathrm{i} \omega_0 t}
\end{eqnarray}
for times $t \in (0,\Delta t)$. If $t$ is not between 0 and $\Delta t$, then the above integral becomes zero. This observation leads us to the spontaneous decay rate
\begin{eqnarray} \label{HC1y}
\Gamma_{\rm free} &=& \frac{e^2 \omega_0^3}{8 \pi^2 c^3 \varepsilon_0 \hbar} \sum_{\lambda = {\sf H}, {\sf V}} 
\int_{\cal S} {\rm d}^2 s \, \big\| {\boldsymbol {\cal D}}_{01} \cdot \boldsymbol{e}_{{\boldsymbol s}\lambda} \big\|^2 \, . ~~
\end{eqnarray}
To further simplify this expression, we now introduce the polar coordinates $\vartheta$ and $\varphi$ with $\vartheta \in [0, \pi]$ and $\varphi \in [0, 2\pi)$ and write the direction vector ${\boldsymbol s}$ and the polarisation vectors ${\boldsymbol e}_{{\boldsymbol{s} \lambda}}$ as
\begin{eqnarray} \label{vectors2}
&& \hspace*{-0.8cm} \left\{ \boldsymbol{s}, \boldsymbol{e}_{\boldsymbol{s} {\sf H}}, \boldsymbol{e}_{\boldsymbol{s} {\sf V}} \right\} \notag \\
&=& \left\{ \begin{pmatrix}\cos \vartheta \\ \cos \varphi \, \sin \vartheta \\ \sin \varphi \, \sin \vartheta \end{pmatrix},
 \begin{pmatrix} 0 \\ ~ \sin\varphi \\ - \cos\varphi \end{pmatrix} ,
 \begin{pmatrix} \sin\vartheta \\ - \cos\varphi\cos\vartheta \\ - \sin\varphi\cos\vartheta \end{pmatrix} \right\} . \notag \\
\end{eqnarray}
One can easily check that the above vectors are pairwise orthonormal. Using Eq.~(\ref{UFx}) and the fact that ${\rm d}^2 s =  {\rm d} \vartheta \, {\rm d} \varphi \, \sin \vartheta$, we find that   
\begin{eqnarray} \label{HC1z}
\Gamma_{\rm free} &=& \frac{e^2 \| {\boldsymbol {\cal D}}_{01} \|^2 \omega_0^3}{8 \pi^2 c^3 \varepsilon_0 \hbar}  
\int_0^{\pi} {\rm d} \vartheta \sin \vartheta \int_0^{2 \pi} {\rm d} \varphi \notag \\
&& \times \left( \cos^2 \varphi +  \sin^2 \varphi  \cos^2 \vartheta \right) \, .
\end{eqnarray}
After performing the $\varphi$ integration, substituting $\xi = \cos \vartheta$ and performing the remaining integration, we obtain the spontaneous decay rate 
\begin{eqnarray} \label{HC1zz}
\Gamma_{\rm free} &=& \frac{e^2 \| {\boldsymbol {\cal D}}_{01} \|^2 \, \omega_0^3}{3 \pi c^3 \varepsilon_0 \hbar} 
\end{eqnarray}
of a single atom in free space which aligns with the quantum optics literature \cite{Hegerfeldt,Stokes}. 

\subsection{A partially transparent mirror interface} \label{sec32}

A partially transparent mirror reduces electric field amplitudes and alters directions of propagation when light arrives at its surface. However, away from the mirror, light still moves along straight lines, as described by Eq.~(\ref{UF}). Taking this into account, placing the mirror into the $x = 0$ plane and the atom at $\boldsymbol{r}_0 = (x_0,0,0)$ with $x_0>0$, as illustrated in Fig.~\ref{figpaperlogo3}, we now find that
\begin{eqnarray} \label{UE1}
~~~ && \hspace*{-1cm} U_{\rm F}(t,0) \boldsymbol{\cal E}^\dagger_{\boldsymbol{s} \lambda}(\boldsymbol{r}_0) |0_{\mathrm{F}} \rangle \notag \\
&=& \left[ \Theta(-s_x) \Theta (x_0 + s_x c t) \, \boldsymbol{\cal E}^\dagger_{\boldsymbol{s} \lambda}(\boldsymbol{r}_0 +\boldsymbol{s} c t) \right.  \notag \\
&& + t_a \,  \Theta(-s_x) \Theta (-(x_0 + s_x c t)) \, \boldsymbol{\cal E}^\dagger_{\boldsymbol{s} \lambda}(\boldsymbol{r}_0 +\boldsymbol{s} c t) \notag \\
 && + r_a \, \Theta(-s_x) \Theta (-(x_0 + s_x c t))\, \boldsymbol{\cal E}^\dagger_{\tilde{\boldsymbol{s}} \lambda}(\tilde{\boldsymbol{r}}_0 + \tilde{\boldsymbol{s}} c t) \notag \\ 
 && \left. + \Theta(s_x) \, \boldsymbol{\cal E}^\dagger_{\boldsymbol{s} \lambda}(\boldsymbol{r}_0 +\boldsymbol{s} c t) \right] |0_{\mathrm{F}} \rangle \, .
\end{eqnarray}
Here the tilde symbol is used to add a minus sign to the $x$ component of a vector. For example, if ${\boldsymbol s} = (s_x,s_y,s_z)$, then $\tilde {\boldsymbol s} = (-s_x,s_y,s_z)$. Hence $\tilde{\mathbf r}_0=(-x_0, 0, 0)$ denotes the position of the mirror image of the atom. In addition, $r_a$ and $t_a$ denote the (complex) reflection and (real) transmission rates of the mirror interface, respectively \cite{AMC}, with
\begin{eqnarray} \label{UE11}
|r_a|^2 + t_a^2 &=& 1 \, .
\end{eqnarray}
In addition, $ \Theta(x)$ has been defined such that
\begin{eqnarray}
\Theta(x)=\left\{\begin{array}{lll}1 & \text { for } & x \geq 0 \, , \\ 0 & \text { for } & x<0 \, . \end{array}\right.
 \end{eqnarray}
The first three terms in Eq.~(\ref{UE1}) describe light which travels away from the atom towards the mirror and is eventually reflected or transmitted (cf.~Fig.~\ref{figpaperlogo3}). Notice that the reflected electric field contributions seem to originate from the mirror image of the atom which is situated a distance of $x_0$ behind the mirror. The final term in Eq.~(\ref{UE1}) describes light which has been emitted away from the mirror surface and propagates as it would in free space. Here we are especially interested in the case where the time it takes light to travel from the atom to the mirror surface is in general much shorter than $\Delta t$. This means, $\Theta \left(-(x_0 + s_x c t) \right) = 1$ and $\Theta \left(x_0 + s_x c t \right) = 0$ for almost all times $t \in (0,\Delta t)$ and 
\begin{eqnarray} \label{UE1z}
U_{\rm F}(t,0) \, \boldsymbol{\cal E}^\dagger_{\boldsymbol{s} \lambda}(\boldsymbol{r}_0) |0_{\mathrm{F}} \rangle 
&=& \left[ t_a \, \Theta(-s_x) \, \boldsymbol{\cal E}^\dagger_{\boldsymbol{s} \lambda}(\boldsymbol{r}_0 +\boldsymbol{s} c t) \right. \notag \\
&& + r_a \, \Theta(-s_x) \, \boldsymbol{\cal E}^\dagger_{\tilde{\boldsymbol{s}} \lambda}(\tilde{\boldsymbol r}_0 + \tilde{\boldsymbol{s}} c t) \notag \\
&& \left. + \Theta(s_x) \, \boldsymbol{\cal E}^\dagger_{\boldsymbol{s} \lambda}(\boldsymbol{r}_0 +\boldsymbol{s} c t) \right] |0_{\mathrm{F}} \rangle ~~~~~
\end{eqnarray}
to a very good approximation.

To calculate the spontaneous decay rate of a single atom in the presence of a partially-transparent mirror interface, we can now proceed as in the previous subsection. Substituting Eqs.~(\ref{E}) and (\ref{UE1z}) into Eq.~(\ref{22}) yields
\begin{widetext}
\begin{eqnarray} \label{22z}
\Gamma_{\rm mir} &=& \frac{e^2}{\hbar^2 \, \Delta t} \sum_{\lambda, \lambda' ={\sf H}, {\sf V}}  
\int_0^{\Delta t} {\rm d}t \int_0^{\Delta t} {\rm d}t' \int_{\mathcal{S}} \mathrm{d}^2 s  \int_{\mathcal{S}} \mathrm{d}^2 s' 
\, {\rm e}^{{\rm i} \omega_0 (t-t')} \, \notag \\
&& \times \langle 0_{\rm F}| \, {\boldsymbol {\cal D}}_{01}  \cdot  
 \Big[ t_a \, \Theta(-s_x') \, \boldsymbol{\cal E}_{\boldsymbol{s}' \lambda'}(\boldsymbol{r}_0 +\boldsymbol{s}' c t') 
+ r_a^* \, \Theta(-s_x') \, \boldsymbol{\cal E}_{\tilde{\boldsymbol{s}}' \lambda'}(\tilde{\boldsymbol r}_0 + \tilde{\boldsymbol{s}}' c t')  
+ \Theta(s_x') \, \boldsymbol{\cal E}_{\boldsymbol{s}' \lambda'}(\boldsymbol{r}_0 +\boldsymbol{s}' c t') \Big] \notag \\
&& \hspace*{0.69cm} \times {\boldsymbol {\cal D}}_{01} \cdot \left[ t_a \, \Theta(-s_x) \, \boldsymbol{\cal E}^\dagger_{\boldsymbol{s} \lambda}(\boldsymbol{r}_0 +\boldsymbol{s} c t) 
 + r_a \, \Theta(-s_x) \, \boldsymbol{\cal E}^\dagger_{\tilde{\boldsymbol{s}} \lambda}(\tilde{\boldsymbol r}_0 + \tilde{\boldsymbol{s}} c t)  
+ \Theta(s_x) \, \boldsymbol{\cal E}^\dagger_{\boldsymbol{s} \lambda}(\boldsymbol{r}_0 +\boldsymbol{s} c t) \right] |0_{\rm F} \rangle \, .
\end{eqnarray}
To evaluate this expression, we employ again Eq.~(\ref{Esl22}) and take into account that state vectors of the electromagnetic field with different final directions of propagation and different polarisations are pairwise orthogonal and therefore do not interfere. Since the atomic dipole moment ${\boldsymbol {\cal D}}_{01}$ is parallel to the mirror surface, we have ${\boldsymbol {\cal D}}_{01} \cdot \boldsymbol{e}_{\boldsymbol{s} \lambda}= {\boldsymbol {\cal D}}_{01} \cdot \tilde {\boldsymbol e}_{\boldsymbol{s} \lambda}$. After performing the $s'$ integration, we therefore find that
\begin{eqnarray} \label{22extra}
\Gamma_{\rm mir} &=& \frac{{\rm i} c e^2 }{8 \pi^2 \varepsilon_0 \hbar \, \Delta t} \sum_{\lambda ={\sf H}, {\sf V}}  
\int_0^{\Delta t} {\rm d}t \int_0^{\Delta t} {\rm d}t' \int_{\mathcal{S}} \mathrm{d}^2 s \, {\rm e}^{{\rm i} \omega_0 (t-t')} \left[ \left( (t_a^2 + |r_a|^2) \Theta(-s_x) + \Theta(s_x) \right) \, \delta^{(3)}( ct - c t' ) \right. \notag \\
&& \left. + r_a^* \, \Theta(s_x) \, \delta^{(3)}( {\boldsymbol s} \cdot (\boldsymbol{r}_0 - \tilde{\boldsymbol{r}}_0) + ct - ct') 
+ r_a \, \Theta(-s_x) \, \delta^{(3)}( \tilde {\boldsymbol s} \cdot (\tilde{\boldsymbol{r}}_0 - \boldsymbol{r}_0) + ct - ct') \right]  \big\| {\boldsymbol {\cal D}}_{01} \cdot \boldsymbol{e}_{{\boldsymbol s}\lambda} \big\|^2 \, . 
\end{eqnarray}
\end{widetext}
Since $(t_a^2 + |r_a|^2) \Theta(-s_x) + \Theta(s_x) = 1$ and ${\boldsymbol s} \cdot \boldsymbol{r}_0 = - \tilde{\boldsymbol s} \cdot \boldsymbol{r}_0 = \tilde {\boldsymbol s} \cdot \tilde {\boldsymbol r}_0$, this equation simplifies to
\newpage 
\begin{eqnarray} \label{22extra2}
\Gamma_{\rm mir} 
&=& \frac{{\rm i} c e^2 }{8 \pi^2 \varepsilon_0 \hbar \, \Delta t} \sum_{\lambda ={\sf H}, {\sf V}}  
\int_0^{\Delta t} {\rm d}t \int_0^{\Delta t} {\rm d}t' \int_{\mathcal{S}} \mathrm{d}^2 s \, {\rm e}^{{\rm i} \omega_0 (t-t')} \notag \\
&& \times \left[ \delta^{(3)}( ct-ct') + {\rm Re} (r_a) \, \delta^{(3)}(2 {\boldsymbol s} \cdot \boldsymbol{r}_0 + ct - ct') \right] \notag \\
&& \times \big\| {\boldsymbol {\cal D}}_{01} \cdot \boldsymbol{e}_{{\boldsymbol s}\lambda} \big\|^2 ,
\end{eqnarray}
in analogy to Eq.~(\ref{22new}). The first term in this equation coincides with the free space decay rate $\Gamma_{\rm free}$. However, the additional terms is due to the far-field interference between reflected light and light which has been emitted away from the mirror.  

\begin{figure}[t]
    \centering
	\includegraphics[width=0.48 \textwidth]{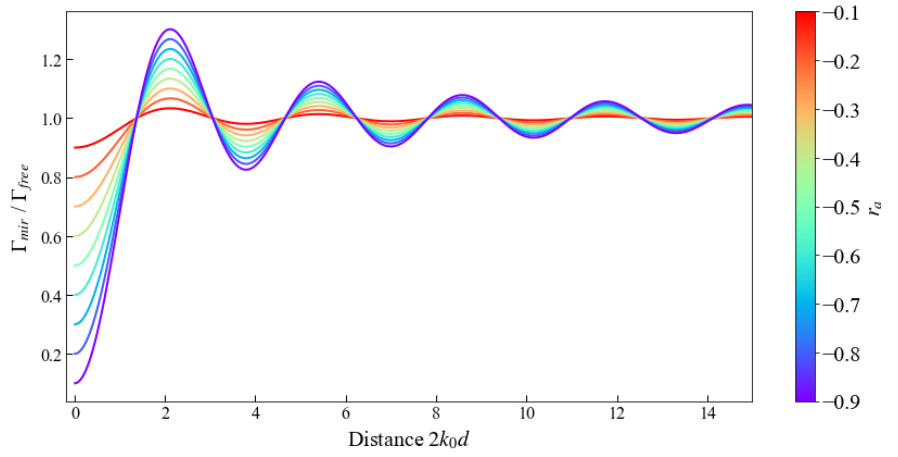}
	\caption{The spontaneous decay rate $\Gamma_{\text {mir }}$ in Eq.~(\ref{E42}) as a function of the emitter-mirror distance $d$ for different reflection coefficients $r_a$. Here we consider dielectric mirrors with real and negative reflection rates $r_a$, since the incoming light accumulates a minus sign upon reflection. The graph highlights the relatively strong dependence of $\Gamma_{\text {mir }}$ on both the emitter-mirror distance and the reflectivity of the surface.}
	\label{figpaperlogo1}
\end{figure}

Hence it is not surprising to see that, by performing the time integrations with the help of Eq.~(\ref{help2}), the spontaneous decay rate $\Gamma_{\rm mir}$ can be shown equal 
\begin{eqnarray} \label{22extra2xxx}
{\Gamma_{\rm mir} \over \Gamma_{\rm free}} &=& 1 + {3 {\rm Re}(r_a) \over 8 \pi} \sum_{\lambda ={\sf H}, {\sf V}}  
\int_{\mathcal{S}} \mathrm{d}^2 s \, {\rm e}^{- 2 {\rm i} k_0  {\boldsymbol s} \cdot \boldsymbol{r}_0} 
 \, \big\| \hat{\boldsymbol {\cal D}}_{01} \cdot \boldsymbol{e}_{{\boldsymbol s}\lambda} \big\|^2 \notag \\
\end{eqnarray}
in units of $\Gamma_{\rm free}$ (cf.~Eq.~(\ref{HC1zz})) and in analogy to Eq.~(\ref{HC1y}). Here the unit vector $\hat{\boldsymbol {\cal D}}_{01}$ is defined such that $\hat{\boldsymbol {\cal D}}_{01} = {\boldsymbol {\cal D}}_{01} /\| {\boldsymbol {\cal D}}_{01} \|$ and $k_0 = \omega_0/c$. When substituting polar coordinates and the vectors in Eq.~(\ref{vectors2}) into this expression, we therefore find that 
\begin{eqnarray} \label{22extra2zzz}
{\Gamma_{\rm mir} \over \Gamma_{\rm free}}
&=&  1 + {3 {\rm Re}(r_a) \over 8 \pi}
\int_0^{\pi} {\rm d} \vartheta \sin \vartheta \int_0^{2 \pi} {\rm d} \varphi \, {\rm e}^{- 2 {\rm i} k_0 d \cos \vartheta} \notag \\
&& \times \left( \cos^2 \varphi +  \sin^2 \varphi  \cos^2 \vartheta \right)  
\end{eqnarray}
with $d = x_0$ denoting the distance between the atom and the mirror surface. Hence 
\begin{eqnarray} \label{E42}
{\Gamma_{\rm mir} \over \Gamma_{\rm free}} 
&=& 1 + {3 {\rm Re}(r_a) \over 2} \notag \\
&& \times \left[ {\sin(2 k_0 d) \over 2 k_0 d} + {\cos(2 k_0 d) \over (2 k_0 d)^2} - {\sin(2 k_0 d) \over (2 k_0 d)^3} \right] ~~~~
\end{eqnarray}
after again carrying out the $\varphi $ and then the $\vartheta$ integration. This result is in good agreement with standard results \cite{Nick,Dorner,Kuraptsev}, thereby verifying the validity of our approach to the derivation of master equations.

\begin{figure}[t]
    \centering
	\includegraphics[width=0.48 \textwidth]{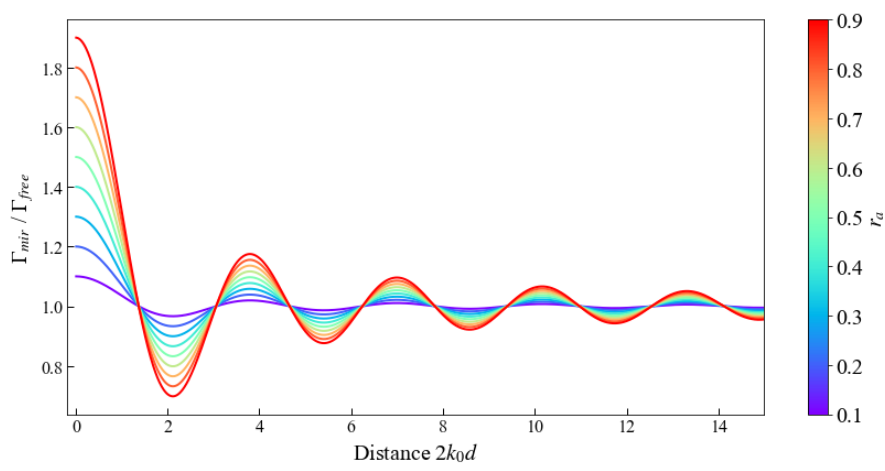}
	\caption{The spontaneous decay rate $\Gamma_{\text {mir }}$ in Eq.~(\ref{E42}) as a function of the emitter-mirror distance $d$ for different reflection coefficients $r_a$. Here we consider a plasmonic mirror surface with positive reflection rates $r_a$, i.e.~the incoming light does not experience a phase shift upon reflection. Now the spontaneous decay rate of the emitter can be twice its free space value.}
	\label{figpaperlogo11}
\end{figure}

\begin{figure*}[t]
\centering
\includegraphics[width=1.4 \columnwidth]{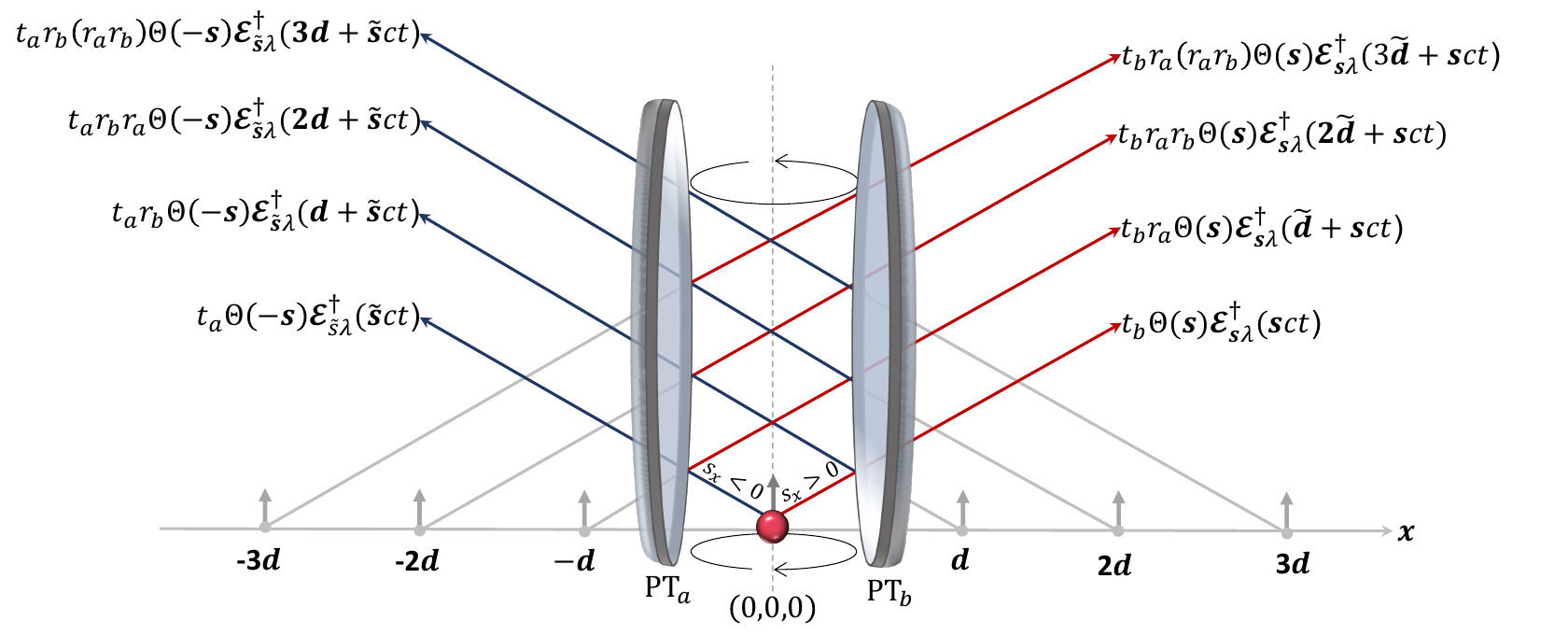}
\caption{[Colour online] Schematic view of an atom placed at the centre of of a cavity which consists of two partially transparent planar mirrors a distance $d$ away from each other. Now $ r_a$, $ r_b$, $t_a$ and $t_b$ denote the (complex) reflection and (real) transmission rates of the interface \cite{AMC}. Compared to the setup in Fig.~\ref{figpaperlogo3}, light emitted from the atom now might get reflected many times before leaving the setup eventually. Now there are infinitely many path which contribute to light emission into the same final direction of propagation and the interference of the outgoing light is strongly enhanced, thereby altering the spontaneous decay rate of the atom.}
\label{figpaperlogo4}
\end{figure*}

As illustrated in Figs.~\ref{figpaperlogo1} and \ref{figpaperlogo11}, for real reflection rates $r_a$, the spontaneous decay rate $\Gamma_{\rm mir}$ of an atom in front of a partially transparent mirror varies between zero and 2$\Gamma_{\rm free}$. For example, when the atom-mirror distance $d$ tends to zero, the light which has been emitted away from the mirror and light which has been reflected by the mirror surface interfere. For highly reflecting mirrors with $r_a =-1$,  $\Gamma_{\rm mir}$ therefore becomes zero (cf.~Fig.~\ref{figpaperlogo1}). The atom cannot loose its energy; its decay channel is blocked. In contrast to this, for $d=0$ and $r_a =1$, the interference is constructive and the emitter looses its initial energy at twice the free space decay rate (cf.~Fig.~\ref{figpaperlogo11}). The above equation also shows that an emitter far away from the mirror interface decays exactly as it would in free space. For more details see for example the discussions in Refs.~\cite{Nick,Dawson,Amos1997,Drexhage}.

\section{The spontaneous decay of a single atom between two planar mirrors} \label{sec4}

There are many similarities between the quantum optics of an atom in front of a highly-reflecting mirror (cf.~Fig.~\ref{figpaperlogo3}) and of an atom between two highly-reflecting mirrors (cf.~Fig.~\ref{figpaperlogo4}). In both systems, the light emerging from the emitter is reflected by the mirror surfaces and the different contributions interfere. This observation suggests that an atom inside a cavity, like an atom near a mirror interface, has only a single spontaneous decay rate $\Gamma_{\rm cav}$ which is in general different from $\Gamma_{\rm free}$. This is in contrast to the standard quantum optics approach to atom-cavity systems in Eq.~(\ref{I2}) which considers two independent decay channels with decay rates $\Gamma$ and $\kappa$. In the following, we calculate the dependence of $\Gamma_{\rm cav}$ on the distance $d$ and the reflection and transmission rates, $r_a$, $r_b$, $t_a$ and $t_b$ of the resonator mirrors using the same methodology as in the previous section. 

For convenience, we place the emitter in the following into the origin of our coordinate system at position ${\boldsymbol r}_0=(0,0,0)$, as illustrated in Fig.~\ref{figpaperlogo4}. Moreover, we place the cavity mirrors, which are parallel to the $x=0$ plane, at positions $x=-d/2$ and $x=d/2$ along the $x$ axis. Proceeding as in the previous Section and considering again sufficiently large times $t$, we now find that
\begin{widetext}
\begin{eqnarray} \label{E45first}
U_{\mathrm{F}}(t, 0) \boldsymbol{\cal E}^\dagger_{{\boldsymbol s} \lambda}(\boldsymbol{r}_0) |0_{\mathrm{F}} \rangle
&=& \Theta(-s_x) \, t_a \left[ \boldsymbol{\cal E}^\dagger_{{\boldsymbol s} \lambda} ({\boldsymbol s} ct)
+ r_a r_b \, \boldsymbol{\cal E}^\dagger_{{\boldsymbol s} \lambda} (2 {\boldsymbol d} + {\boldsymbol s} ct) 
+ r_a^2 r_b^2 \, \boldsymbol{\cal E}^\dagger_{{\boldsymbol s} \lambda} (4 {\boldsymbol d} + {\boldsymbol s} ct) 
+ \ldots \, \right]  |0_{\mathrm{F}} \rangle \notag \\
&& + \Theta(-s_x) \, r_a t_b \left[ \boldsymbol{\cal E}^\dagger_{\tilde{\boldsymbol s} \lambda} (- {\boldsymbol d} + \tilde {\boldsymbol s} ct) 
+ r_a r_b \, \boldsymbol{\cal E}^\dagger_{\tilde{\boldsymbol s} \lambda} (- 3 {\boldsymbol d} + \tilde {\boldsymbol s} ct) 
+ r_a^2 r_b^2 \, \boldsymbol{\cal E}^\dagger_{\tilde{\boldsymbol s} \lambda} (- 5 {\boldsymbol d} + \tilde {\boldsymbol s} ct) 
+ \ldots \, \right]  |0_{\mathrm{F}} \rangle \notag \\
&& + \Theta(s_x) \, t_b \left[ \boldsymbol{\cal E}^\dagger_{{\boldsymbol s} \lambda} ({\boldsymbol s} ct)
+ r_a r_b \, \boldsymbol{\cal E}^\dagger_{{\boldsymbol s} \lambda} (- 2 {\boldsymbol d} + {\boldsymbol s} ct) 
+ r_a^2 r_b^2 \, \boldsymbol{\cal E}^\dagger_{{\boldsymbol s} \lambda} (- 4 {\boldsymbol d} + {\boldsymbol s} ct) 
+ \ldots \, \right]  |0_{\mathrm{F}} \rangle \notag \\
&& +  \Theta(s_x) \, r_b t_a \left[ \boldsymbol{\cal E}^\dagger_{\tilde{\boldsymbol s} \lambda} ({\boldsymbol d} + \tilde {\boldsymbol s} ct) 
+ r_a r_b \, \boldsymbol{\cal E}^\dagger_{\tilde{\boldsymbol s} \lambda} (3 {\boldsymbol d} + \tilde {\boldsymbol s} ct) 
+ r_a^2 r_b^2 \, \boldsymbol{\cal E}^\dagger_{\tilde{\boldsymbol s} \lambda} (5 {\boldsymbol d} + \tilde {\boldsymbol s} ct) 
+ \ldots \, \right]  |0_{\mathrm{F}} \rangle 
\end{eqnarray}
in the absence of light-matter interactions and with ${\boldsymbol d} = (d,0,0)$ and $d>0$. The first line in this equation describes light which has been created at $t=0$, travels to the left and eventually leaves the cavity on the left hand side. Analogously, the second line describes light which also travels left initially but eventually leaves the cavity on the right. Analogously, the last two lines describe light which travels initially to the right. After combining the above terms into geometric sums, Eq.~(\ref{E45first}) simplifies to
\begin{eqnarray} \label{E45}
U_{\mathrm{F}}(t, 0) \boldsymbol{\cal E}^\dagger_{\boldsymbol{s} \lambda}\left(\boldsymbol{r}_0\right) |0_{\mathrm{F}}\rangle
&= & \sum_{n=0}^\infty \left( r_a r_b\right)^n 
\left[  t_a \, \Theta(-s_x) \, \boldsymbol{\cal E}^\dagger_{\boldsymbol{s} \lambda}(2 n {\boldsymbol d}+\boldsymbol{s} c t)
+ t_b \,  \Theta (s_x) \, \boldsymbol{\cal E}^\dagger_{\boldsymbol{s} \lambda}(-2 n {\boldsymbol d}+\boldsymbol{s} c t) \right. \notag \\ 
&& \left. + r_a t_b \, \Theta(-s_x) \, \boldsymbol{\cal E}^\dagger_{\tilde{\boldsymbol{s}} \lambda}(- (2 n+1) {{\boldsymbol d}}+\tilde{\boldsymbol{s}} c t)  
+ r_b t_a \, \Theta (s_x) \, \boldsymbol{\cal E}^\dagger_{\tilde{\boldsymbol{s}} \lambda}((2n+1) {{\boldsymbol d}}+\tilde{\boldsymbol{s}} c t) \right] |0_{\mathrm{F}}\rangle 
\end{eqnarray}
in analogy to Eq.~(\ref{UE1z}). The above state is the (unnormalised) state of a single photon with polarisation $\lambda$ at time $t$ which has been emitted by the atom into the ${\boldsymbol s}$ direction at $t=0$. 

Proceeding as in the previous section and substituting Eq.~(\ref{E45}) into Eq.~(\ref{22}), we therefore find that the spontaneous decay rate $\Gamma_{\rm cav}$ of the atom inside the cavity equals
\begin{eqnarray} \label{extra}
\Gamma_{\rm cav} &=& \frac{e^2}{\hbar^2 \, \Delta t} \sum_{n,n'=0}^\infty 
\sum_{\lambda, \lambda'={\sf H}, {\sf V}} \int_0^{\Delta t} {\rm d} t \int_0^{\Delta t} \mathrm{d} t' \int_{\mathcal{S}} \mathrm{d}^2 s \int_{\mathcal{S}} \mathrm{d}^2 s' \, \left(r_a r_b\right)^{n} \left(r_a^* r_b^* \right)^{n'} \, {\rm e}^{{\rm i} \omega_0 (t-t')}  \notag \\
&& \times \langle 0_{\rm F}| \, {\boldsymbol {\cal D}}_{01}  \cdot  
\Big[ \, t_a \, \Theta(-s_x') \, \boldsymbol{\cal E}_{\boldsymbol{s}' \lambda'}(2 n' {\boldsymbol d}+\boldsymbol{s}' c t')
+ t_b \,  \Theta (s_x') \, \boldsymbol{\cal E}_{\boldsymbol{s}' \lambda}(-2 n' {\boldsymbol d}+\boldsymbol{s}' c t') \notag \\ 
&& \hspace*{2.1cm} + r_a^* t_b \, \Theta(-s_x') \, \boldsymbol{\cal E}_{\tilde{\boldsymbol{s}}' \lambda'}(- (2 n'+1) {{\boldsymbol d}}+\tilde{\boldsymbol{s}}' c t')  
+ r_b^* t_a \, \Theta (s_x') \, \boldsymbol{\cal E}_{\tilde{\boldsymbol{s}}' \lambda'}((2n'+1) {{\boldsymbol d}}+\tilde{\boldsymbol{s}}' c t') \Big] \notag \\
&& \hspace*{0.69cm} \times {\boldsymbol {\cal D}}_{01} \cdot 
\left[ \, t_a \, \Theta(-s_x) \, \boldsymbol{\cal E}^\dagger_{\boldsymbol{s} \lambda}(2 n {\boldsymbol d}+\boldsymbol{s} c t)
+ t_b \,  \Theta (s_x) \, \boldsymbol{\cal E}^\dagger_{\boldsymbol{s} \lambda}(-2 n {\boldsymbol d}+\boldsymbol{s} c t) \right. \notag \\ 
&& \hspace*{2.1cm} \left. + r_a t_b \, \Theta(-s_x) \, \boldsymbol{\cal E}^\dagger_{\tilde{\boldsymbol{s}} \lambda}(- (2 n+1) {{\boldsymbol d}}+\tilde{\boldsymbol{s}} c t)  
+ r_b t_a \, \Theta (s_x) \, \boldsymbol{\cal E}^\dagger_{\tilde{\boldsymbol{s}} \lambda}((2n+1) {{\boldsymbol d}}+\tilde{\boldsymbol{s}} c t) \right]  |0_{\rm F} \rangle \, .
\end{eqnarray}
It is not surprising that the above expression has many similarities with $\Gamma_{\rm mir}$ in Eq.~(\ref{22z}); the same interference effects contribute to the changes of the spontaneous decay rate compared to the free-space case. The only difference here is that we now have to take into account not only one but infinitely many reflections. Next we notice that states vectors of the electromagnetic field with different final directions of propagation and different polarisations are pairwise orthogonal. This observation allows us to sum over one of the polarisations and to perform the ${\boldsymbol s}'$ integration. Substituting Eq.~(\ref{Esl22}) into Eq.~(\ref{extra}) and considering again the real dipole moment ${\boldsymbol {\cal D}}_{01}$ in Eq.~(\ref{UFx}) with $ \| {\boldsymbol {\cal D}}_ {01} \cdot \boldsymbol {e} _ { {\boldsymbol s} \lambda } \| ^ { 2 }  =  \| {\boldsymbol {\cal D}}_ {01} \cdot \boldsymbol {e} _ { \tilde {\boldsymbol s} \lambda } \| ^ { 2 } $, the spontaneous decay rate $\Gamma_{\rm cav}$ becomes
\begin{eqnarray} \label{extrabold}
\Gamma_{\rm cav} &=& \frac{{\rm i} c e^2 }{8 \pi^2 \varepsilon_0 \hbar \, \Delta t} \sum_{n,n'=0}^\infty 
\sum_{\lambda ={\sf H}, {\sf V}} \int_0^{\Delta t} {\rm d} t \int_0^{\Delta t} \mathrm{d} t' \int_{\mathcal{S}} \mathrm{d}^2 s 
\, \left(r_a r_b\right)^{n} \left(r_a^* r_b^* \right)^{n'}   
\, {\rm e}^{{\rm i} \omega_0 (t-t')} \, \big\| {\boldsymbol {\cal D}}_{01} \cdot \boldsymbol{e}_{{\boldsymbol s}\lambda} \big\|^2 \notag \\
&& \times \left[  
t_a^2 \, \Theta(-s_x) \, \delta^{(3)}( 2 (n-n') \boldsymbol{s} \cdot {\boldsymbol d} + c t - c t' ) 
+ t_b^2 \, \Theta(s_x) \, \delta^{(3)}( - 2 (n-n') \boldsymbol{s} \cdot {\boldsymbol d} + c t - c t' ) \right. \notag \\
&& \hspace*{0.25cm} + |r_a|^2 t_b^2 \, \Theta(-s_x) \, \delta^{(3)}( -2 (n-n') \tilde{\boldsymbol{s}} \cdot {\boldsymbol d} + c t - c t' ) 
+ |r_b|^2 t_a^2 \, \Theta(s_x) \, \delta^{(3)}( 2 (n-n') \tilde{\boldsymbol{s}} \cdot {\boldsymbol d} + c t - c t' ) \notag \\
&& \hspace*{0.25cm} + r_b t_a^2 \, \Theta (s_x) \, \delta^{(3)}( (2n - 2n' +1) \tilde{\boldsymbol{s}} \cdot {\boldsymbol d} + c t - c t' ) 
+ r_b^* t_a^2 \, \Theta (-s_x) \, \delta^{(3)}( (2n - 2n' -1) \boldsymbol{s} \cdot {\boldsymbol d} + c t - c t' ) \notag \\
&& \left. \hspace*{0.25cm} + r_a t_b^2 \, \Theta (-s_x) \, \delta^{(3)}( -(2n - 2n' + 1) \tilde{\boldsymbol{s}} \cdot {\boldsymbol d} + c t - c t' ) + r_a^* t_b^2 \, \Theta (s_x) \, \delta^{(3)}( -(2n - 2n' - 1) \boldsymbol{s} \cdot {\boldsymbol d} + c t - c t' )
\right] , ~~~~
\end{eqnarray}
\end{widetext}
in analogy to Eq.~(\ref{22extra}).

For simplicity, we restrict ourselves in the following to symmetric cavities with real reflection rates. In this case, the reflection and transmission rates of both mirrors are the same and 
\begin{eqnarray} \label{extrabold2}
r_a = r_b = r_{\rm mir} ~~ {\rm and} ~~ t_a = t_b = t_{\rm mir} \, .
\end{eqnarray}
In addition, we take into account that $\Theta(-s_x) + \Theta(s_x) = 1$ and that we can exchange $n$ and $n'$ without changing the right hand side of Eq.~(\ref{extrabold3}).  Moreover, we notice that $\tilde{\boldsymbol{s}} \cdot {\boldsymbol d} = - \boldsymbol{s} \cdot {\boldsymbol d}$. Subsequently we find that
\begin{eqnarray} \label{extrabold3}
\Gamma_{\rm cav} &=& \frac{{\rm i} c e^2 }{8 \pi^2 \varepsilon_0 \hbar \, \Delta t} \sum_{n,n'=0}^\infty 
\sum_{\lambda ={\sf H}, {\sf V}} \int_0^{\Delta t} {\rm d} t \int_0^{\Delta t} \mathrm{d} t' \int_{\mathcal{S}} \mathrm{d}^2 s ~~~ \notag \\
&& \times r_{\rm mir}^{2(n+n')} \, t_{\rm mir}^2 \, {\rm e}^{{\rm i} \omega_0 (t-t')} \, \big\| {\boldsymbol {\cal D}}_{01} \cdot \boldsymbol{e}_{{\boldsymbol s}\lambda} \big\|^2 \notag \\
&& \times \left[ \left(1 + r_{\rm mir}^2 \right) \, \delta^{(3)}( 2 (n-n') \boldsymbol{s} \cdot {\boldsymbol d} + c t - c t' ) \right. \notag \\
&& + r_{\rm mir} \, \delta^{(3)}( (2n - 2n' +1) \boldsymbol{s} \cdot {\boldsymbol d} + c t - c t' ) \notag \\
&& \left. + r_{\rm mir} \, \delta^{(3)}( (2n - 2n' - 1) \boldsymbol{s} \cdot {\boldsymbol d} + c t - c t' ) \right] 
\end{eqnarray}
which has many similarities with $\Gamma_{\rm mir}$ in Eq.~(\ref{22extra2}). To perform the time integrations, we assume in the following that $|\boldsymbol{s} \cdot {\boldsymbol d}|$ is much smaller than $ct$ for almost all times $t \in (0,\Delta t)$. In this case, we can evaluate the above time integrals again with the help of Eq.~(\ref{help2}). Subsequently, we find that  
\begin{eqnarray} \label{extrabold5}
{\Gamma_{\rm cav} \over \Gamma_{\rm free}} &=& \frac{3}{8 \pi} \sum_{n,n'=0}^\infty 
\sum_{\lambda ={\sf H}, {\sf V}} \int_{\mathcal{S}} \mathrm{d}^2 s \, r_{\rm mir} ^{2(n+n')} \, t_{\rm mir}^2 \, \big\| \hat{\boldsymbol {\cal D}}_{01} \cdot \boldsymbol{e}_{{\boldsymbol s}\lambda} \big\|^2 \notag \\
&& \times {\rm e}^{- 2{\rm i} (n-n') k_0 \boldsymbol{s} \cdot {\boldsymbol d}} \, \left\| 1 + r_{\rm mir} \, {\rm e}^{- {\rm i} k_0 \boldsymbol{s} \cdot {\boldsymbol d}} \right\|^2
\end{eqnarray}
to a very good approximation, in analogy to Eq.~(\ref{22extra2xxx}). Next we have a closer look at the implications of the above result for subwavelength cavities and optical cavities with highly-reflecting mirrors.

\subsection{Subwavelength cavities}

To study the spontaneous emission from an emitter inside a subwavelength cavity, we now perform the summations in Eq.~(\ref{extrabold5}) using the geometric series formulae for complex numbers to show that 
\begin{eqnarray} \label{extrabold30}
{\Gamma_{\rm cav} \over \Gamma_{\rm free}} &=& \frac{3}{8 \pi}
\sum_{\lambda ={\sf H}, {\sf V}} \int_{\mathcal{S}} \mathrm{d}^2 s  \, t_{\rm mir}^2 \, 
\big\| \hat{\boldsymbol {\cal D}}_{01} \cdot \boldsymbol{e}_{{\boldsymbol s}\lambda} \big\|^2 \notag \\
&& \times \left\| {1 + r_{\rm mir} \, {\rm e}^{- {\rm i} k_0 \boldsymbol{s} \cdot {\boldsymbol d}} \over
1 - r_{\rm mir}^2 \, {\rm e}^{- 2 {\rm i} k_0 \boldsymbol{s} \cdot {\boldsymbol d}}}  \right\|^2 \, .
\end{eqnarray}
In the case of a subwavelength cavity, the distance $d$ between the resonator mirrors is much smaller than the optical wavelength $\lambda_0$ of the atomic transition and $k_0d \ll 1$.  Taking this into account and substituting polar coordinates and the vectors in Eq.~(\ref{vectors2}) into this equation and taking into account that $ t_{\rm mir}^2 +  |r_{\rm mir}|^2 = 1$, we find that
\begin{eqnarray} \label{extrabold20}
\lim_{d \to 0} {\Gamma_{\rm cav} \over \Gamma_{\rm free}} 
&=& \frac{3}{8 \pi} \cdot {1 + r_{\rm mir} \over 1 - r_{\rm mir}} \int_0^{\pi} {\rm d} \vartheta \, \sin \vartheta \notag \\
&& \times \int_0^{2 \pi} {\rm d} \varphi \left( \cos^2 \varphi +  \sin^2 \varphi  \cos^2 \vartheta \right) ~~~
\end{eqnarray}
in zeroth order in $k_0 d$. This equation leads us to
\begin{eqnarray} \label{extrabold21}
\lim_{d \to 0} {\Gamma_{\rm cav} \over \Gamma_{\rm free}} 
&=& {1 + r_{\rm mir} \over 1 - r_{\rm mir}} \, .
\end{eqnarray}
Starting again from Eq.~(\ref{extrabold30}), using again polar coordinates but keeping terms proportional to $(k_0 d)^2$, we moreover find that 
\begin{eqnarray} \label{extrabold24}
{\Gamma_{\rm cav} \over \Gamma_{\rm free}} 
&=& {1 + r_{\rm mir} \over 1 - r_{\rm mir}} \left[ 1 - {2 \over 5} \cdot { r_{\rm mir} \over (1- r_{\rm mir} )^2} \,  (k_0 d)^2 \right]  ~~~~
\end{eqnarray}
up to second order in $k_0 d$. As one might expect, the spontaneous decay rate $\Gamma_{\rm cav}$ of an emitter inside a subwavelength cavity depends strongly on the reflection and transmission rates of the resonator mirrors. However, it depends only relatively weakly on the distance $d$ of the resonator mirrors.

\begin{figure}[t]
\centering
\includegraphics[width=0.99 \columnwidth]{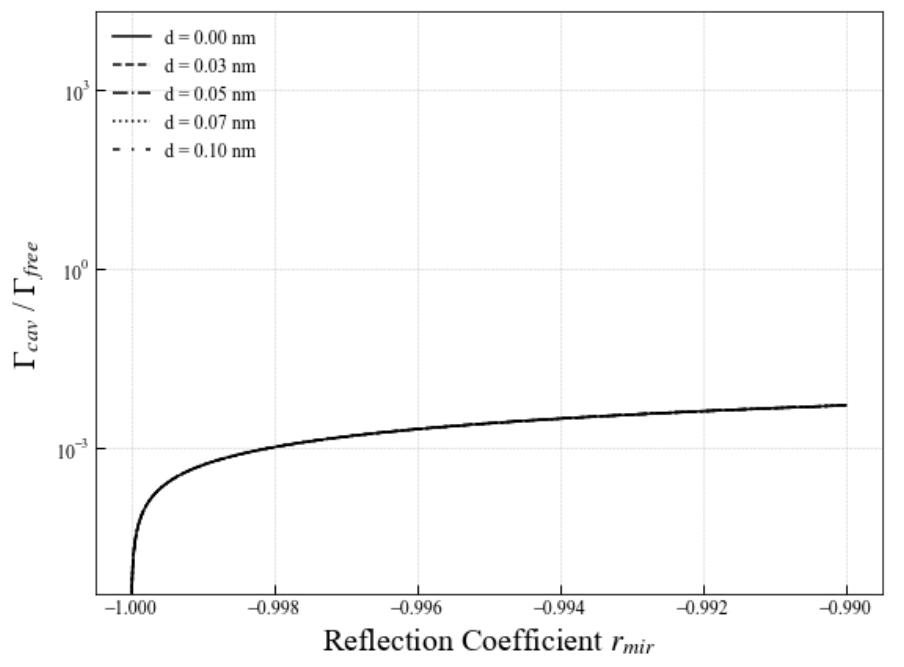}
\caption{[Colour online] The spontaneous decay rate $\Gamma_{\rm cav}$ in Eq.~(\ref{extrabold24}) of a single emitter inside a subwavelength cavity with highly-reflecting dielectric mirrors as a function of the reflection rate $r_{\rm mir}$ for different 
mirror distances. In this case, $r_{\rm mir}$ is real and negative \cite{Hecht,AMC}. When $r_{\rm mir} = -1$, the spontaneous decay rate $\Gamma_{\rm cav}$ becomes zero. This is not surprising, since the light coming from the emitter cannot leave the resonator in this case.}
\label{figpaperlogo5}
\end{figure}

\begin{figure}[t]
\centering
\includegraphics[width=0.99 \columnwidth]{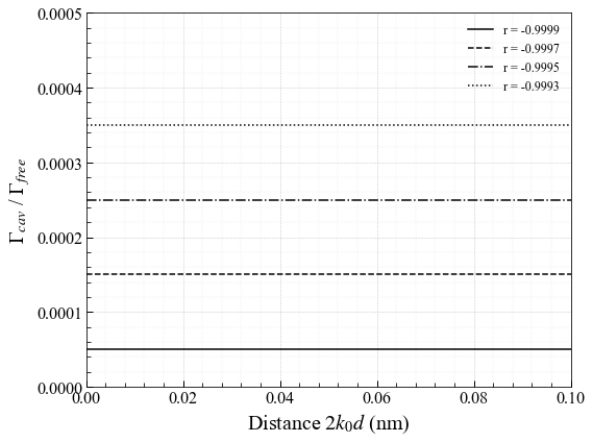}
\caption{[Colour online] The spontaneous decay rate $\Gamma_{\rm cav}$ in Eq.~(\ref{extrabold24}) of an emitter inside a subwavelength cavity with dielectric mirrors as a function of the mirror distance $d$ for different negative reflection rates $r_{\rm mir}$. As long as $k_0 d \ll 1$, there is essentially no difference between Eqs.~(\ref{extrabold21}) and (\ref{extrabold24}). In all the cases that we consider here, $\Gamma_{\rm cav}$ is much smaller than the free space decay rate $\Gamma_{\rm free}$.}
\label{figpaperlogo55}
\end{figure}

For photonic crystal subwavelength cavities \cite{Yablonovitch1987,Jelena2,Lohdahl2004,Krishnamoorthy2010}, the resonator mirrors are formed by a dielectric material and have negative reflection rates $r_{\rm mir}$, since electric field amplitudes must remain continuous on the mirror surface in this case. This continuity condition only applies when the electric field amplitude of the incoming light accumulates a minus sign upon reflection \cite{Hecht,AMC}. For photonic crystal cavities, highly reflecting mirrors therefore correspond to the case $r_{\rm mir} = -1$. Hence, as illustrated in Figs.~\ref{figpaperlogo5} and \ref{figpaperlogo55}, the decay rate $\Gamma_{\rm cav}$ of an emitter inside a very narrow photonic crystal cavity with highly-reflecting mirrors is in general much smaller than its free space decay rate. Moreover, we find that the spontaneous decay rate $\Gamma_{\rm cav}$ of an atom inside a weakly-reflecting subwavelength cavity $(r_{\rm mir} \to 0)$ coincides with the free space decay rate $\Gamma_{\rm free}$, as one would expect due to the similarities between both situations.

\begin{figure}[t]
\centering
\includegraphics[width=0.99 \columnwidth]{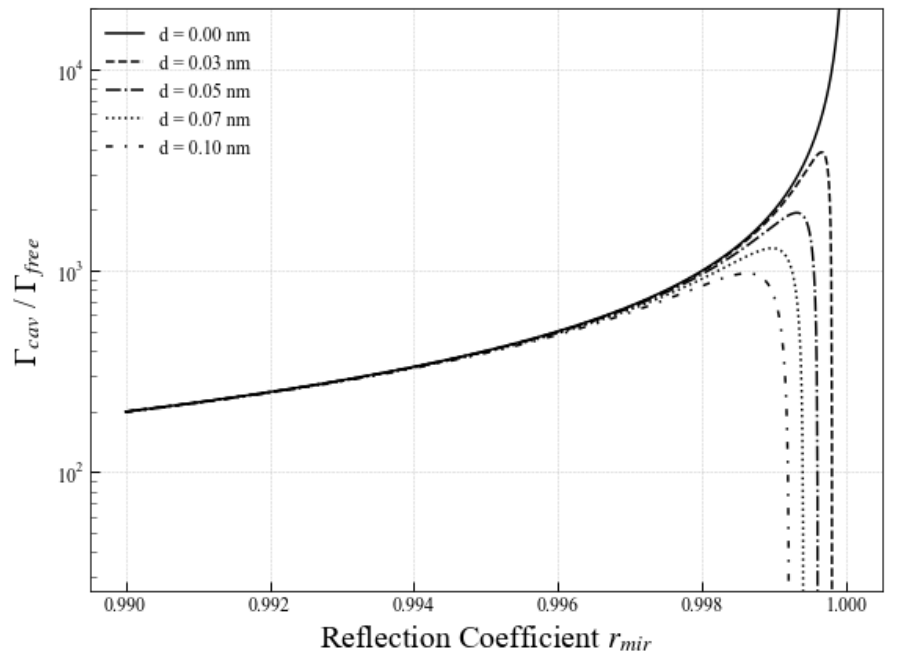}
\caption{[Colour online] The spontaneous decay rate $\Gamma_{\rm cav}$ in Eq.~(\ref{extrabold24}) of an emitter inside a subwavelength cavity with plasmonic mirrors as a function of mirror reflection rates $r_{\rm mir}$ for different mirror distances $d$. This figure confirms that the spontaneous decay rate $\Gamma_{\rm cav}$ of an emitter inside a plasmonic nanocavity can be significantly larger than its free space decay rate $\Gamma_{\rm free}$.}
\label{figpaperlogo6}
\end{figure}

\begin{figure}[t]
\centering
\includegraphics[width=0.99 \columnwidth]{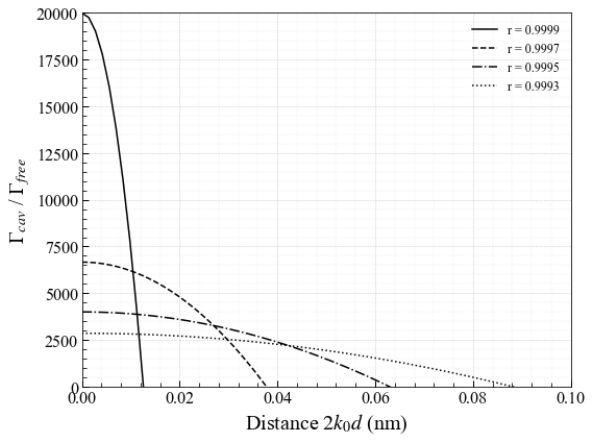}
\caption{[Colour online] The spontaneous decay rate $\Gamma_{\rm cav}$ in Eq.~(\ref{extrabold24}) of a subwavelength cavity with highly-reflecting plasmonic mirrors \cite{Baumberg} as a function of the mirror distance $d$ for different positive reflection rates $r_{\rm mir}$. The figure illustrates that, in the cases considered here, $\Gamma_{\rm cav}$ can be much larger than $\Gamma_{\rm free}$. However, this enhancement is in general relatively short range. This observation is in good agreement with the decay rates $\Gamma_{\rm cav}$ of planar optical cavities which we obtained in Section \ref{sec43}. These are always close to $\Gamma_{\rm free}$ when the emitter-mirror distance is of the same order of magnitude or larger than the wavelength $\lambda_0$ of the emitted light.}
\label{figpaperlogo66}
\end{figure}

Next we have a closer look at light emission by atoms in subwavelength cavities with metallic mirrors or mirrors made up of other metasurfaces \cite{complex,Barreda2021,Liu2025}. For example, in the experiments of Baumberg's group, individual atoms are made to fall through a small gap between tiny gold-coated spheres \cite{Baumberg}. In this case, surface currents which could be due to the roughness of the mirrors need to be taken into account. As a result of plasmonic couplings, the light coming from the atoms no longer accumulates a minus sign upon reflection by a mirror surface. In general, the reflection rate of metallic mirrors has been found to be complex \cite{Fox}. In subwavelength cavity 
with highly-reflecting mirrors, where the electromagnetic field maintains its phase upon reflection, $r_{\rm mir}$ is positive and close to one. Eq.~(\ref{extrabold24}) therefore shows that the spontaneous decay rate $\Gamma_{\rm cav}$ can be much larger than $\Gamma_{\rm free}$ in this case, as illustrated in Figs.~\ref{figpaperlogo6}. However, Fig.~\ref{figpaperlogo66} shows that the enhancement of the emission rate of the atom is relatively short range. It requires very small mirror distances $d$ which seems to be in good agreement with experimental findings \cite{Pelton,Baumberg,Baumberg2,Russell2012,Li2024}. 

\subsection{Planar optical cavities with highly-reflecting mirrors} \label{sec43}

To study the dynamics of an emitter which has been placed into the centre of a planar optical cavity, where $d$ is at least comparable to the transition wavelength $\lambda_0$ of the atom or much larger, we start again from Eq.~(\ref{extrabold5}). However, this time, we do not perform the summations over $n$ and $n'$. Instead, we introduce polar coordinates and substitute the vectors in Eq.~(\ref{vectors2}) into Eq.~(\ref{extrabold5}) to show that 
\begin{eqnarray} \label{extrabold6}
{\Gamma_{\rm cav} \over \Gamma_{\rm free}} &=& \frac{3}{8 \pi} \sum_{n,n'=0}^\infty 
\int_0^{\pi} {\rm d} \vartheta \, \sin \vartheta \int_0^{2 \pi} {\rm d} \varphi \, r_{\rm mir} ^{2(n+n')} \, t_{\rm mir}^2 \notag \\
&& \times \left( \cos^2 \varphi +  \sin^2 \varphi  \cos^2 \vartheta \right) {\rm e}^{- 2{\rm i} (n-n') k_0 d \cos \vartheta} \notag \\
&& \times \left\| 1 + r_{\rm mir} \, {\rm e}^{- {\rm i} k_0 d \cos \vartheta} \right \|^2 
\end{eqnarray}
without any approximations. Now we can perform first the $\varphi$ and then the $\vartheta$ integration. After replacing the variable $n'$ by a new variable $m=n-n'$, our calculations lead us to
\begin{eqnarray} \label{extrabold7}
{\Gamma_{\rm cav} \over \Gamma_{\rm free}} 
&=& \frac{3}{2} \sum_{n=0}^\infty \sum_{m=-\infty}^n r_{\rm mir} ^{4n-2m} \,  t_{\rm mir}^2  \left[ \left(1 + r_{\rm mir}^2 \right) f (2m k_0 d) \right. \notag \\
&& \left. + r_{\rm mir} \, f((2m - 1) k_0 d) + r_{\rm mir} \, f((2m + 1) k_0 d) \right] \notag \\
\end{eqnarray}
with the function $f(x)$ given by
\begin{eqnarray} \label{extrabold8}
f(x) &=& {\sin x \over x} + {\cos x \over x^2} - {\sin x \over x^3} \, .
\end{eqnarray}
This function assumes its maximum value at $x=0$ and $f(0) = 2/3$. However, when $x$ is non-zero, the terms contributing to $f(x)$ tend to zero very rapidly. Hence, only the $m=0$ terms contributes in this case to the spontaneous decay rate $\Gamma_{\rm cav}$, in the case which we consider here where $k_0 d \gg 0$. Taking this into account, Eq.~(\ref{extrabold7}) simplifies to
\begin{eqnarray} \label{extrabold9}
{\Gamma_{\rm cav} \over \Gamma_{\rm free}} 
&=& \sum_{n=0}^\infty r_{\rm mir} ^{4n} \,  t_{\rm mir}^2 \left(1 + r_{\rm mir}^2 \right) \notag \\
&=& {(1- r_{\rm mir}^2) (1+ r_{\rm mir}^2) \over 1-r_{\rm mir}^4} \notag \\
&=& 1 \, .
\end{eqnarray}
This equation shows that the spontaneous decay rate $\Gamma_{\rm cav}$ of an atom at the centre of a planar optical cavity coincides essentially with its free space decay rate $\Gamma_{\rm free}$. Notice that this result holds in the limit $k_0 d \gg 1$, and is therefore independent of the mirror separation $d$, the mirror reflectivity $r_{\mathrm{mir}}$, and the phase shifts upon reflection, provided the atom is placed at the centre of the resonator. This limit corresponds to standard optical cavities in which $d \gg \lambda_0$.

The above result might seem surprising, if one assumes that the atom couples to a single standing wave cavity field mode whose photons leak out through the resonator mirrors. Taking this point of view, one would expect that the decay of the atom happens at a rate which depends predominantly on the reflection and transmission rates of the resonator mirrors. In contrast to this, our calculations show that the emitter should be considered a local source of electromagnetic radiation in the form of travelling waves which bounce back and further between the resonator mirrors. The light generated by the emitter interferes in the same way, light interferes when travelling though a Fabry-Perot cavity \cite{Abeer}. However, as we can see from Eqs.~(\ref{extrabold7}) and (\ref{extrabold8}), this interference becomes negligible, if the mirror distance $d$ is much larger than the wavelength $\lambda_0$ of the emitted light. An atom inside a cavity is essentially an emitter in the presence of {\em two} mirror surfaces and hence evolves similarly to an atom in the presence of a single mirror which we discussed in Section \ref{sec32}.

\section{Conclusions} \label{sec5}


The so-called cooperativity parameter $C$ is widely accepted as a measure for how fast an atom-cavity system looses its initial excitation. Within the framework of the Jaynes-Cummings model \cite{theory,theory2,theory3,theory4}, this parameter can be written as $C = g^2/\kappa \Gamma_{\rm free}$ where $g$ is the atom-field interaction constant in Eq.~(\ref{I1}), while $\kappa$ and $\Gamma_{\rm free}$ denote the cavity and the emitter free-space decay rates, respectively. In principle, the parameter $C$ can assume any value. For example, it seems that it is possible to enter the so-called strong coupling regime with $C \gg 1$ by simply increasing the reflectivity of the resonator mirrors which increases the lifetime $1/\kappa$ of photons inside the resonator. Moreover, the Purcell effect \cite{Purcell1946} suggests that reducing the distance of the cavity mirrors increases the atom-field coupling constant $g$ by reducing the mode volume of the cavity photons. Entering the strong coupling regime is important, since many atom-cavity quantum computing schemes require $C$ to be at least two or even three orders of magnitude larger than one \cite{Metz}. It is worth noting that the strong-coupling regime with $C \gg 1$ has indeed been demonstrated experimentally in atom--cavity systems~(cf.~e.g.~Refs.~\cite{Kuhn,Reichel}).


However, achieving this regime in standard optical cavities invariably requires specially engineered configurations, such as carefully shaped mirror geometries designed to promote photon re-absorption by the emitter~\cite{Kuhn, Reichel}. Our calculations show that $C \gg 1$ cannot be achieved with conventional Fabry--P\'{e}rot resonators with planar mirrors and $k_0d \gg 1$. For these, $\Gamma_{\rm cav} = \Gamma_{\rm free}$ (cf.~Eq.~(\ref{extrabold9})) and the cooperativity parameter $C$ seems to be close to one. To show this, this paper takes an alternative approach to the modelling of light propagation in atom-cavity systems. Usually it is assumed that the presence of the cavity mirrors imposes boundary conditions on the quantised electromagnetic field, thereby reducing its Hilbert space to a discrete set of standing wave modes. The available frequencies $\omega_{\rm cav}$ of these modes depend only on the distance $d$ of the resonator mirrors. In contrast to this, this paper assumes that the presence of a mirror surface does not alter the structure of the quantised electromagnetic field. It only alters the dynamics of the photons \cite{Jake,AMC}. 


The only exception that we found are atom-cavity systems with planar mirrors for which the distance $d$ of the resonator mirrors is much smaller than the wavelength $\lambda_0$ of the emitted light. In the case of positive mirror reflection rates $r_{\rm mir}$, the spontaneous decay rate $\Gamma_{\rm cav}$ in Eq.~(\ref{extrabold24}) can become much larger than its free space decay rate $\Gamma_{\rm free}$ (cf.~Fig.~\ref{figpaperlogo6}). This enhancement requires that the resonator mirrors are metasurfaces which do not change the sign of electric field amplitudes upon reflection, as it would be the case for light reflection by dielectric media \cite{Hecht,AMC}. Such metasurfaces exist \cite{complex,Barreda2021,Liu2025}. For example, experiments have shown that metallic mirrors with some surface roughness have complex reflection rates \cite{Fox} due to the possibility of currents within the mirror surface. However, in general, an emitter placed at the center of two planar mirrors decays at a rate similar to its free space decay rate $\Gamma_{\rm free}$ (cf.~Eq.~(\ref{extrabold9})), independent of the reflection and transmission rates and of the distance of the resonator mirrors.

One way of enhancing atom-cavity interactions while reducing photon emission is to consider subwavelength cavities with dielectric mirrors which have negative reflection rates \cite{AMC}. Alternatively, one could try to avoid one of the assumptions which we made in our calculations for simplicity. These are:
\begin{enumerate}
\item Here we only consider cavities with planar mirrors which is normally not the case in atom-cavity experiments \cite{Kuhn,Reichel}. Usually, the mirrors are shaped such that they refocus the light back onto the emitter to encourage re-absorption by the source. 

\item Here we neglected the re-absorption of light due to the relatively small size of an atom compared to the wavelength $\lambda_0$ of the emitted light (cf.~Eq.~(\ref{E45})). For example, increasing the size of the emitter might alter interference effects and might increase re-absorption.

\item Another assumption made in this paper is that the atom is placed exactly at the centre of the resonator (cf.~Fig.~\ref{figpaperlogo4}). For an atom at a position $x_0 \neq 0$, light coming from the emitter evolves differently and Eq.~(\ref{E45}) for the complex electric field vector $U_{\mathrm{F}}(t, 0) \boldsymbol{\cal E}^\dagger_{\boldsymbol{s} \lambda}\left(\boldsymbol{r}_0\right) |0_{\mathrm{F}}\rangle$ needs to be adjusted accordingly. Notice that this field vector is all that is needed to calculate the spontaneous decay rate $\Gamma_{\mathrm{cav}}$ of the emitter inside a planar cavity (cf.~Eq.~(\ref{22})). Incorporating this position dependence is a well-defined extension of the present framework and constitutes a natural direction for future work. Moving the emitter to a different position is likely to alter the interference within the resonator and subsequently $\Gamma_{\mathrm{cav}}$.

\item Our calculations also assume that the reflection rates $r_a$ and $r_b$ on the insides of the resonator mirrors are both real and both the same, i.e.~$r_a = r_b = r_{\rm mir}$. Breaking this symmetry is also likely to change the amount of interference inside the resonator. However, our discussion in Section \ref{sec32} suggests that changing the balance between $r_a$ and $r_b$ is unlikely to strongly affect the result in Eq.~(\ref{extrabold9}), as long as the distance of the atom from the cavity mirrors remains relatively large. 

\item Finally, our calculations assume that the mirror reflectivity $r_{\mathrm{mir}}$ is independent of the angle of incidence. In the case of a dipole moment vectors $\boldsymbol{D}_{01}$ parallel to the mirror surface (cf.~Eq.~(\ref{UFx})), this approximation is justified since the dominant contribution to $\Gamma_{\mathrm{cav}}$ originates from light emitted at near-normal incidence ($\vartheta \approx 0$). The angular variation of the reflectivity is therefore negligible to a very good approximation, for both metallic and dielectric mirrors. For other orientations of the dipole moment $\boldsymbol{D}_{01}$, similar arguments can be used to replace the generally angle-dependent mirror reflection rate by a single most relevant value. We therefore expect that our conclusions remain qualitatively valid also in more general scenarios.
\end{enumerate}
We believe that the relatively simple analysis that we present already provides some interesting new insights. For example, it explains why it is so hard to design optical cavities which operate in the so-called strong coupling regime. It also explains some dramatic differences in the emission dynamics of atoms in metallic subwavelength cavities and in photonic bandgap materials. \\[0.5cm]
{\em Acknowledgements.} This research was funded by the Government of the Kingdom of Saudi Arabia. Moreover, this work was supported in part by the Oxford Quantum Technology Hub NQIT (grant number EP/M013243/1).

\end{document}